\documentclass{aa}  

\usepackage[normalem]{ulem}
\usepackage{natbib}
\usepackage{graphicx,textpos}
\usepackage{booktabs} 
\usepackage{txfonts}
\usepackage{xcolor}
\usepackage{hyperref}
\usepackage{rotating}
\usepackage[utf8]{inputenc}
\usepackage{gensymb}
\DeclareUnicodeCharacter{00A0}{~}
\bibpunct{(}{)}{;}{a}{}{,}
\newcommand{\referee}[1]{\textcolor{black}{#1}}
\newcommand{\Ana}[1]{\textcolor{black}{#1}}

\newcommand{\kms}{km\,${\rm s}^{-1}$}
\newcommand{\Msun}{M$_{\odot}$}

\newcommand{\Teff}{$T_{\rm eff}$\,}
\newcommand{\logg}{$\log\,g$\,}
\newcommand{\FeH}{[Fe/H]\,}
\newcommand{\alphaM}{[$\alpha$/M]\,}
\newcommand{\vsini}{{$\varv\,\sin i$\,}}
\newcommand{\vmac}{$\varv_\mathrm{macro}$\,}
\newcommand{\vmic}{$\varv_\mathrm{micro}$\,}
\newcommand{\ispec}{\texttt{iSpec}}

\begin{document}

\title{Stellar nucleosynthesis in the era of large surveys:\\
S-process-polluted binaries in GALAH DR4}

\author{A. Escorza\inst{1,2} \and
S. Vitali\inst{3,1,2} \and
D. Godoy-Rivera\inst{1,2} \and
S. Shetye\inst{4} \and
H. Van Winckel\inst{4} \and
G. Bustos\inst{5} \and
S. Goriely\inst{6} \and
L. Siess\inst{6} \and
M. Abdul-Masih\inst{1,2} \and
T. Masseron\inst{1,2} \and
D. A. García-Hernández\inst{1,2} \and
A. Ardern-Arentsen\inst{7} \and
P. Jofré\inst{8} \and
S. Van Eck\inst{6}
}

\institute{Instituto de Astrofísica de Canarias, C. Vía Láctea, s/n, 38205 La Laguna, Santa Cruz de Tenerife, Spain\\
\email{aescorza@iac.es}
         \and
Universidad de La Laguna, Dpto. Astrofísica, Av. Astrofísico Francisco Sánchez, 38206 La Laguna, Santa Cruz de Tenerife, Spain.
         \and
INAF - Osservatorio Astrofisico di Torino, Strada Osservatorio 20, I-10025 Pino Torinese (TO), Italy.
         \and
Instituut voor Sterrenkunde, KU Leuven, Celestijnenlaan 200D bus 2401, Leuven, 3001, Belgium
         \and
Universidad Internacional de Valencia (VIU), C/Pintor Sorolla 21, E-46002 Valencia, Spain
         \and
Institute of Astronomy and Astrophysics (IAA), Université libre de Bruxelles (ULB), CP 226, Boulevard du Triomphe, B-1050 Bruxelles, Belgium
         \and
Institute of Astronomy, University of Cambridge, Madingley Road, Cambridge CB3 0HA, UK
         \and
Instituto de Estudios Astrofísicos, Facultad de Ingeniería y Ciencias, Universidad Diego Portales, Av. Ejército Libertador 441, Santiago, Chile}

\date{\today}

\abstract
{\Ana{Binary interactions during the asymptotic giant branch phase can lead to the formation of chemically peculiar stars such as Ba-, CH-, and CEMP-s stars with overabundances of s-process elements as well as carbon.  Only a few hundred of these stars have been subject to detailed chemical and/or dynamical studies, typically in non-homogeneous samples.}}
{We compile a \Ana{systematic} sample of s-process-polluted candidates using the fourth data release of the large spectroscopic survey GALAH. \Ana{We also compare the chemical and orbital properties of these candidates with those of confirmed s-process-polluted stars to gain stronger evidence of their nature.}}
{\Ana{GALAH DR4 uses neural networks and automatic spectral analysis methods as well as data} of a lower spectral resolution than is normally used to characterise \Ana{these} objects. \Ana{We therefore built a validation sample of s-rich stars before querying the survey, for which we obtained VLT/UVES and Mercator/HERMES high-resolution (>80\,000) spectra and analysed them with traditional methods}. We compared our stellar parameters and individual abundances with those of the survey and used this validation to define the thresholds that a star in GALAH DR4 must pass to be flagged as a \Ana{good s-process-rich candidate. Based on our comparisons with the high-resolution complementary spectra, we defined thresholds on five quantities: $\rm{[s/Fe]} > 0.40$~dex, $\rm{[Y/Fe]} > 0.22$~dex, $\rm{[Zr/Fe]} > 0.26$~dex, $\rm{[Ba/Fe]} > 0.19$~dex, and $\rm{[La/Fe]} > 0.26$~dex.}}
{We identified \referee{1059} stars in GALAH DR4 that are good candidates to be s-process-polluted stars, covering a broad parameter space \Ana{(4000~K~<~\Teff~<~6000~K, 0.5~dex~<~\logg~<~5.0~dex, -2.0~dex~<~[Fe/H]~<~0.5~dex)}. They share many similarities with \Ana{the samples of confirmed s-rich stars, especially their ratios of heavy over light s-elements}, which strengthens our confidence in \Ana{the purity of the sample}. When we cross-matched the sample with the literature, we found that only 7\% of the candidates have measured orbital periods and eccentricities. This limits a full comparison with confirmed \Ana{Ba and related star samples for now. However, their binary fraction is, as expected, higher than the one we found for the full GALAH DR4 catalogue.}}
{\Ana{Our sample of candidates is almost five times larger than the number of currently confirmed polluted stars. This and the fact that it has been homogeneously treated by the GALAH survey opens very interesting avenues to compare nucleosynthesis models as well as binary evolution models.}}

  
\maketitle

\section{Introduction}\label{intro}

About half of the elements heavier than iron are synthesised by the slow neutron-capture (s-) process of nucleosynthesis \citep[e.g.][]{Burbidge57, Clayton61, Kappeler11}, and the main astrophysical site where this process occurs is the interior of asymptotic giant branch (AGB) stars \citep[e.g.][]{Lugaro03long, Karakas10}. When AGB stars evolve in binary systems, their nucleosynthesis \Ana{products} can be transferred to an unevolved companion during a mass-transfer episode. Several families of s-process-enriched objects that are too young to have produced these elements themselves populate the literature on low- to intermediate-mass binary interaction products.

At slightly subsolar metallicity \citep[\Ana{$-0.5<\rm{[Fe/H]}<0.0$;}][]{deCastro16, Escorza17}, barium (Ba) stars \citep{BidelmanKeenan51} are the prototypical example of AGB-binary interaction products. Their binarity and the white dwarf nature of their companions are well accepted \citep[e.g.][]{McClure80, McClure84, Udry98, Jorissen98, Bohm-Vitense00, Pourbaix2000, Jorissen19, Escorza19, EscorzaDeRosa23}. In the more metal-poor regime, CH stars and carbon-enhanced metal-poor stars with s-element overabundances (CEMP-s stars) are closely related to Ba stars, and they are often considered as their Population II analogues. They show similar s-element overabundances together with strong CH molecular bands \citep[e.g.][]{Keenan42, McClureWoodsworth90, Jorissen16}. \Ana{Moreover, there also exist extrinsic S-type stars, which, like intrinsic S stars, display ZrO and TiO bands. However, because they lack technetium, they are understood not to be currently undergoing s-process nucleosynthesis or third dredge-up, and their heavy-element enrichment is attributed to past accretion in a binary system \citep[e.g.][]{IbenRenzini83, Jorissen98, VanEck99, Jorissen19, Shetye18, Shetye20, Shetye21}.} Finally, s-process-polluted stars also appear in clusters among the populations of blue and yellow stragglers \citep[e.g.][]{Milliman15, Nine24} and are thought to have the same binary origin as the other families listed above.

It is commonly stated that in the galactic field, about 1\% of the red giants are Ba stars \citep{MacConnell72, NorthDuquennoy91}, and at lower metallicity, this fraction is expected to be much higher \citep[e.g.][]{Abate15} because the efficiency of the s-process is higher \citep[e.g.][]{Karakas14}. However, these occurrence rates have not been reviewed recently using high-resolution spectroscopy or large and homogeneous samples, and this revision is one of the goals of this work. During the past decade, we have witnessed a flourishing of large spectroscopic surveys of different sizes over different areas of the sky and with different spectral resolutions and wavelength ranges that are producing an incredible amount of data. Among these, the GALactic Archaeology with HERMES (GALAH) survey \citep{DeSilva2015, 2021Buder, 2025Buder} is especially interesting in the search for s-process-polluted stars. GALAH DR4 \citep{2025Buder} provides 1,085,520 spectra obtained with the High Efficiency and Resolution Multi-Element Spectrograph (HERMES) at the Anglo-Australian Telescope \citep{Sheinis2015} and reduced as described by \cite{Kos2017} and \cite{Zwitter2021}. Their spectra cover the wavelength range between 4713\AA\ and 7887\AA\ with a spectral resolution of 28\,000. Additionally, they also provide the atmospheric parameters for 917,588 stars and the individual abundances of 32 chemical elements. These 32 species include eight s-process elements \Ana{(Rb, Sr, Y, Zr, Ba, La, Ce, Nd)}, which is unprecedented for these types of surveys.

The previous GALAH data release \citep[GALAH DR3;][]{2021Buder} already made significant contributions to neutron-capture nucleosynthesis \citep[e.g.][]{Aguado21, Matsuno21, Griffith22, Horta22} and to the search for barium stars \citep[e.g.][]{Rekhi24, Rekhi26, Levine26}. However, the improvements from GALAH DR3 to GALAH DR4 are significant and particularly important for determining stellar parameters and individual abundances. In particular, the signal-to-noise ratios (S/N) have increased because the survey repeated the observations of previous pointings \Ana{(see Fig. 4 in \citealt{2025Buder})}. Additionally, GALAH DR4 uses a broader wavelength range than DR3 \Ana{because their new method is less computationally expensive and allows them to use the full spectrum. Finally, it} includes non-spectroscopic information and an improved normalisation method. \Ana{The results presented in GALAH DR4 are based on neural networks trained on SME \citep[Spectroscopy Made Easy; ][]{SME1996, SME2017} synthetic spectra, which led to} a more accurate comparison of synthetic and observed spectra, and thus, to a more accurate stellar label optimisation \citep{2025Buder}. \Ana{These improvements in the method are demonstrated by the validation of} their results against literature values (mainly the \textit{Gaia} FGK benchmark stars; \citealt{Jofre14, Jofre15, Jofre18, Heiter15}), especially in the precision achieved for multiple elements, including the s-process elements Y, La, Ce, Nd, and Sm (see Sect. 6.2 in \citealt{2025Buder} for further details). \Ana{While this paper was in preparation, \cite{Yang26} published the first catalogue of Ba stars using GALAH DR4. While the authors relied on more than one s-process elements (La and Ba), their threshold for flagging a polluted star is too optimistic, as we show in Sect.~\ref{sec:bastars}.}

We build a \Ana{sample} of s-process-polluted stars across the GALAH parameter space using their DR4 catalogue. \Ana{We place this new sample in the context of already confirmed polluted stars and analyse their chemical and orbital properties. Moreover, using this large sample of homogeneously analysed candidates, we also review the occurrence rate of s-process-polluted stars at different metallicities within the GALAH DR4 range.} To do this, we rely on several s-process elements, not only on barium, which is known to be a particularly problematic element to measure because some spectral lines saturate \citep[e.g.][]{Karinkuzhi21}, because it suffers from non-local thermodynamical equilibrium (NLTE) effects \citep{Short06, Korotin15, Gallagher20}, \Ana{and because its isotopic composition is unknown}. Sect.~\ref{sec:GALAHprep} describes our element choices among the options provided by GALAH DR4. Additionally, since the GALAH neural networks are not trained on very high abundances and they might underperform for chemically peculiar stars \citep{2025Buder}, we built a high-resolution (HR) validation sample to perform our own catalogue validation in the abundance range of interest. \Ana{This has not been done by any of the previous works searching for s-process-enhanced stars in GALAH data releases. }This HR sample and the observations are described in Sect.~\ref{sec:HRsample}, and \Ana{the spectral analysis performed on the sample is described} in Sect.~\ref{sec:specanalysis}. In Sect.~\ref{sec:comparison} we compare our HR results with those of GALAH DR4 to establish abundance thresholds that can help us produce a reliable list of s-process-polluted candidates for which we prioritise purity over completeness, and in Sect.~\ref{sec:bastars} we describe this final list. Finally, in Sect. \ref{sec:discussion} we place our candidates in the context of well-known s-process-polluted stars and discuss binarity in the sample, since all these polluted candidates are expected to be in long-period binaries. Our main conclusions are then summarised in Sect.~\ref{sec:summary}.

\begin{figure*}[t]
    \centering
    \includegraphics[width=\textwidth]{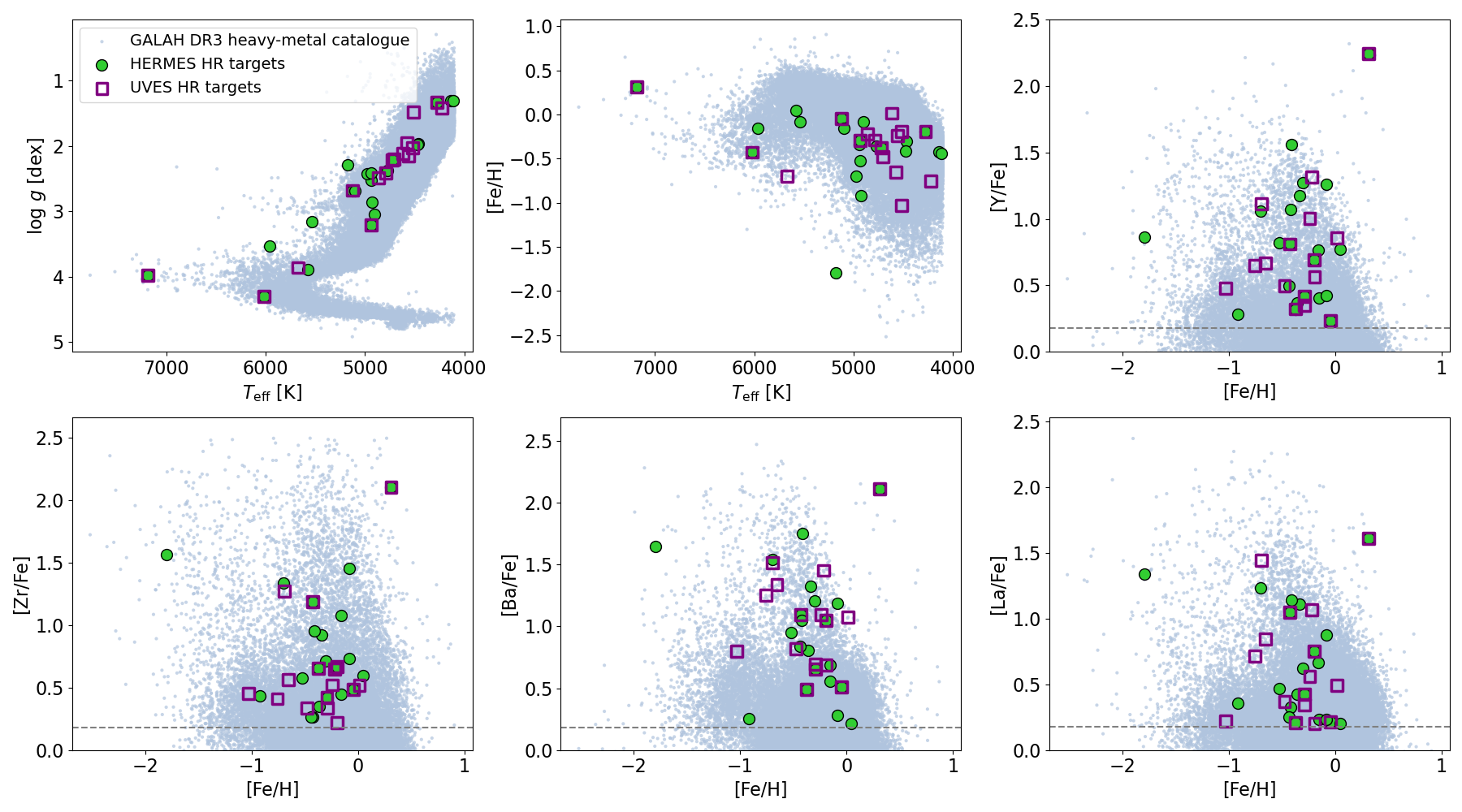}
    \caption{High-resolution validation sample observed with HERMES (green circles) and UVES (purple squares) plotted over a clean GALAH DR3 catalogue \citep{2021Buder} containing stars for which the spectroscopic analysis was reliable and good-quality abundances of Y, Zr, Ba, and La \Ana{were obtained}. We plot the HR sample against the GALAH DR3 catalogue and not the GALAH DR4 equivalent because the HR target selection was performed before GALAH DR4 was published, and we aim to show our efforts to represent the parameter space. \referee{A copy of this figure using GALAH DR4 is included in Appendix~\ref{appendix:Fig1DR4} for illustration.}}
    \label{Fig:HRtargets}
\end{figure*}

\section{Heavy metal abundances in GALAH DR4}\label{sec:GALAHprep}

The main GALAH DR4 catalogue contains data for 917,588 stars, but since our science case depends on high-quality stellar parameters and chemical abundances, we imposed several quality cuts, including those suggested in the GALAH documentation. First, we only considered stars with flags indicating that the presented spectroscopic analysis is reliable (flag\_sp = 0 and flag\_sp\_fit = 0) and that the data reduction was successful (flag\_red = 0). Then, we discarded stars with a signal-to-noise ratio below 30 and kept those with a reliable metallicity flag (flag\_fe\_h = 0). These quality cuts reduced the catalogue to \referee{53\% of its original size, to 484,105 stars}. We refer to this subset as the “spectroscopic catalogue”.

The next step was to ensure the quality of the chemical abundances in which we were interested. We wished to identify s-process-rich stars, but we would rather not rely on a single element. GALAH DR4 provides individual abundances for four light s-process (ls) elements (Rb, Sr, Y, and Zr) and four heavy s-process (hs) elements (Ba, La, Ce, and Nd). However, it is very restrictive to search for stars with an abundance measurement and a good-quality flag for all these elements. For example, \referee{only 21\%} of the stars in the "spectroscopic catalogue" have a high-quality Sr abundance and \referee{25\% have a high-quality Ce abundance}. After carefully validating the effect of selecting more or fewer elements or selecting one set of elements or a different one, we decided to work with Y, Zr, Ba, and La. This selection includes two ls and two hs elements and allowed us to evaluate the abundances ratios of the two groups without relying on single elements. Additionally, these \Ana{four} elements have several good lines \Ana{in the GALAH wavelength range}, so their abundances do not rely on single lines. This facilitated our comparison with high-resolution data. Finally, ensuring that our subsample had good-quality abundances of these four elements (flag\_x\_fe = 0 with x = y, zr, ba and la in each case) reduced the sample size less when compared to other possible combinations. We refer to this subset as the “heavy-metal catalogue”, and it contains \referee{159,464 stars, $\sim33$\% of the stars in the "spectroscopic catalogue" and $\sim17$\% of the stars in GALAH DR4.}

\section{High-resolution validation sample}\label{sec:HRsample}

\subsection{Target list}\label{sec:HRtargets}
In order to confirm that the GALAH DR4 catalogue can be \Ana{used to confidently identify s-process-enhanced stars}, we built a high-resolution validation sample of 29 stars to compare our \Ana{independently obtained abundances} with the results of the survey. With the HR sample, we tried to cover the parameter space of GALAH \Ana{in terms of effective temperature (\Teff, between 4000~K and about 7000~K), surface gravity (\logg, between 1.0~dex and 5.0~dex), metallicity (\FeH, between -2.0~dex and 0.5~dex)}, and the abundances of the four heavy metals we chose ([Y/Fe], [Zr/Fe], [Ba/Fe], and [La/Fe]\footnote{$\left[\frac{\rm X}{\rm Fe}\right] = \log \left(\frac{N_{\rm X}}{N_{\rm H}}\right)_* - \log \left(\frac{N_{\rm X}}{N_{\rm H}}\right)_{\odot} - \left[\frac{\rm Fe}{\rm H}\right]$\\
with $\left[\frac{\rm Fe}{\rm H}\right] = \log \left(\frac{N_{\rm Fe}}{N_{\rm H}}\right)_* - \log \left(\frac{N_{\rm Fe}}{N_{\rm H}}\right)_{\odot}$\\
and $N_{\rm X}$ the number of atoms of
element X.} \Ana{ between 0.2~dex and 2.0~dex)}.
We note that the data collection for this project started before GALAH DR4 was published, and our target selection was based on GALAH DR3 \citep{2021Buder}.

\Ana{To build the validation sample,} we planned two observing campaigns, one in the northern hemisphere with the HERMES spectrograph (High-Efficiency and high-resolution Mercator {\'e}chelle Spectrograph; R$\sim$85\,000; \citealt{Raskin11, Raskin14}) on the Mercator Telescope (1.2 m mirror), and another one in the South with the UVES spectrograph (Ultraviolet and Visual {\'E}chelle Spectrograph; R$\sim$80\,000; \citealt{UVES}) on the Very Large Telescope (8.2 m mirror). Given the smaller size of the Mercator Telescope,  with HERMES we simply observed the brightest (Vmag < 11) northern targets of the “heavy-metal catalogue” (see Sect. \ref{sec:GALAHprep}) in GALAH DR3. Most of these data were obtained under programme 55-Mercator4/24B (PI: A. Escorza) of the Spanish observing time in the Canarian observatories, although these targets are now included in evolved-binary monitoring programs. Then, a UVES proposal (111.24UZ; PI: A. Escorza) was built to cover gaps in the parameter space. Figure \ref{Fig:HRtargets} shows the full validation HR sample over the GALAH DR3 version of the “heavy-metal catalogue”, \referee{and a copy of this figure using GALAH DR4 is included in Appendix~\ref{appendix:Fig1DR4}} for illustration. Table \ref{table:HRtargets} lists the HR targets, some relevant information about them, and the characteristics of the observations obtained. We note that three stars were observed with both HERMES and UVES.

After a first quick visual inspection of the data and to ensure a homogeneous analysis for all targets, we discarded from the sample \Ana{five targets that were cooler (\Teff~<~3500~K) than reported by GALAH and} whose optical spectra were dominated by molecular bands. These stars are likely AGB stars and would require a different analysis method than we present in Sect.~\ref{sec:specanalysis}. Because of this, they are not included in Table \ref{table:HRtargets} or in Fig. \ref{Fig:HRtargets} and will be presented in a future publication, together with some additional targets that we plan to add to this “cold HR sample” (Shetye et al. in prep).

\subsection{Observations}\label{sec:HRobs}

The HERMES sample was observed using the high-resolution fibre mode, which has a resolving power of $R = 85\,000$ and covers the wavelength range from 3900\,\AA\ to 9000\,\AA. The 2D spectral images were reduced using the dedicated HERMES pipeline \citep{Raskin11, Raskin14}, which accounts for flat-fielding, bias subtraction, cosmic-ray removal, order merging, and wavelength calibration to produce the final 1D spectra with which we worked. 

The UVES sample was observed using the dichroic mode with central wavelengths \Ana{390\,nm and 564\,nm}. In this way, the spectra cover the wavelength range from 3320\,\AA\ to 6680\,\AA\ with a gap between approximately 4520\,\AA\ and 4625\,\AA. We used the 0.5” slit, which provides a resolving power of about $80\,000$. The data were automatically reduced by ESO using the standard UVES data reduction pipeline \citep{Ballester2000}.

Table \ref{table:HRtargets} lists the exposure times used for each observation, as well as the S/N values provided by each data reduction pipeline. For HERMES, it corresponds with the S/N at order 65 (approximately at 550~nm). For UVES, we report separately the S/N for each arm at their centre wavelength (390~nm and 564~nm).

\begin{table*}[t]
\caption{Stellar parameters and individual abundances obtained for our HR sample as described in Sect.~\ref{sec:specanalysis}.}
\label{table:HRresults}
\centering
\begin{small}
\begin{tabular}{lcccccccc}
\toprule
Star ID & Instrument & \Teff & \logg & \FeH & [Y/Fe] & [Zr/Fe] & [Ba/Fe] & [La/Fe]\\
\midrule
BD-00\,1784 & HERMES & 6100 $\pm$ 40 & 4.14 $\pm$ 0.04 & -0.51 $\pm$ 0.03 & 0.75 $\pm$ 0.09 & 0.37 $\pm$ 0.17 & 1.25 $\pm$ 0.07 & 0.99 $\pm$ 0.08\\
 & UVES & 6000 $\pm$ 30 & 3.93 $\pm$ 0.04 & -0.51 $\pm$ 0.03 & 0.74 $\pm$ 0.04 & 0.37 $\pm$ 0.07 & 1.17 $\pm$ 0.07 & 0.92 $\pm$ 0.05\\
BD-01\,2730 & HERMES & \referee{4693 $\pm$ 15} & 2.26 $\pm$ 0.03 & -0.45 $\pm$ 0.06 & 0.23 $\pm$ 0.11 & 0.24 $\pm$ 0.07 & 0.69 $\pm$ 0.04 & 0.46 $\pm$ 0.08\\
BD-03\,2082 & UVES & \referee{4638 $\pm$ 10} & 2.32 $\pm$ 0.03 & -0.31 $\pm$ 0.11 & 0.1 $\pm$ 0.3 & 0.33 $\pm$ 0.11 & 0.70 $\pm$ 0.06 & 0.35 $\pm$ 0.17\\
BD-03\,3480 & HERMES & 5970 $\pm$ 30 & 3.65 $\pm$ 0.03 & -0.20 $\pm$ 0.05 & 0.50 $\pm$ 0.09 & 0.1 $\pm$ 0.4 & 0.80 $\pm$ 0.05 & 0.23 $\pm$ 0.05\\
BD-17\,4712 & UVES & \referee{4475 $\pm$ 12} & 1.59 $\pm$ 0.03 & -0.77 $\pm$ 0.04 & 0.54 $\pm$ 0.07 & 0.54 $\pm$ 0.05 & 1.27 $\pm$ 0.06 & 0.98 $\pm$ 0.11\\
CD-25\,3581 & UVES & \referee{4486 $\pm$ 7} & 2.05 $\pm$ 0.02 & -0.24 $\pm$ 0.03 & 0.48 $\pm$ 0.13 & 0.47 $\pm$ 0.03 & 1.06 $\pm$ 0.03 & 0.67 $\pm$ 0.08\\
CD-43\,12242 & UVES & \referee{4725 $\pm$ 12} & 2.28 $\pm$ 0.03 & -0.37 $\pm$ 0.04 & 0.13 $\pm$ 0.10 & 0.16 $\pm$ 0.05 & 0.53 $\pm$ 0.04 & 0.38 $\pm$ 0.06\\
CD-46\,7169 & UVES & 4700 $\pm$ 20 & 2.12 $\pm$ 0.05 & -0.51 $\pm$ 0.04 & 0.30 $\pm$ 0.08 & 0.26 $\pm$ 0.06 & 0.79 $\pm$ 0.07 & 0.39 $\pm$ 0.06\\
HD\,118367 & UVES & \referee{4518 $\pm$ 8} & 1.24 $\pm$ 0.02 & -1.20 $\pm$ 0.04 & 0.31 $\pm$ 0.08 & 0.63 $\pm$ 0.08 & 0.50 $\pm$ 0.05 & 0.31 $\pm$ 0.06\\
HD\,170752 & UVES & \referee{4131 $\pm$ 10} & 1.00 $\pm$ 0.03 & -0.85 $\pm$ 0.05 & 0.30 $\pm$ 0.16 & 0.47 $\pm$ 0.06 & 1.20 $\pm$ 0.07 & 0.86 $\pm$ 0.09\\
HD\,25584 & HERMES & \referee{4901 $\pm$ 15} & 2.58 $\pm$ 0.04 & -0.25 $\pm$ 0.04 & 0.00 $\pm$ 0.11 & 0.10 $\pm$ 0.07 & 0.63 $\pm$ 0.04 & 0.16 $\pm$ 0.06\\
 & UVES & \referee{4832 $\pm$ 17} & 2.42 $\pm$ 0.05 & -0.30 $\pm$ 0.04 & -0.06 $\pm$ 0.14 & 0.02 $\pm$ 0.07 & 0.47 $\pm$ 0.05 & 0.10 $\pm$ 0.07\\
HD\,285405 & HERMES & \referee{4866 $\pm$ 15} & 2.15 $\pm$ 0.04 & -0.43 $\pm$ 0.05 & 0.92 $\pm$ 0.09 & 0.97 $\pm$ 0.06 & 1.67 $\pm$ 0.03 & 1.11 $\pm$ 0.12\\
HD\,285923 & HERMES & \referee{4838 $\pm$ 14} & 2.80 $\pm$ 0.03 & -0.18 $\pm$ 0.04 & 0.90 $\pm$ 0.09 & 0.82 $\pm$ 0.05 & 1.29 $\pm$ 0.05 & 0.84 $\pm$ 0.13\\
HD\,285933 & HERMES & \referee{4138 $\pm$ 13} & 1.33 $\pm$ 0.04 & -0.38 $\pm$ 0.09 & 0.3 $\pm$ 0.3 & 0.37 $\pm$ 0.09 & 0.78 $\pm$ 0.06 & 0.48 $\pm$ 0.13\\
HD\,295095 & HERMES & \referee{4928 $\pm$ 12} & 3.09 $\pm$ 0.02 & -0.32 $\pm$ 0.05 & 0.42 $\pm$ 0.09 & 0.39 $\pm$ 0.06 & 0.80 $\pm$ 0.04 & 0.41 $\pm$ 0.05\\
 & UVES & \referee{4887 $\pm$ 15} & 2.96 $\pm$ 0.03 & -0.34 $\pm$ 0.05 & 0.35 $\pm$ 0.08 & 0.33 $\pm$ 0.06 & 0.60 $\pm$ 0.04 & 0.36 $\pm$ 0.06\\
HD\,315809 & UVES & 5450 $\pm$ 30 & 3.02 $\pm$ 0.06 & -0.93 $\pm$ 0.06 & 0.95 $\pm$ 0.07 & 0.8 $\pm$ 0.3 & 1.59 $\pm$ 0.04 & 1.32 $\pm$ 0.07\\
HD\,325083 & UVES & \referee{4585 $\pm$ 9} & 2.30 $\pm$ 0.03 & -0.14 $\pm$ 0.05 & 0.31 $\pm$ 0.15 & 0.27 $\pm$ 0.05 & 0.49 $\pm$ 0.07 & 0.35 $\pm$ 0.06\\
TYC\,169-2026-1 & HERMES & 5180 $\pm$ 20 & 2.83 $\pm$ 0.05 & -0.19 $\pm$ 0.05 & 0.03 $\pm$ 0.13 & 0.15 $\pm$ 0.12 & 0.53 $\pm$ 0.06 & 0.22 $\pm$ 0.10\\
TYC\,179-2074-1 & HERMES & \referee{4432 $\pm$ 14} & 1.77 $\pm$ 0.06 & -0.29 $\pm$ 0.07 & 0.68 $\pm$ 0.15 & 0.67 $\pm$ 0.08 & 1.22 $\pm$ 0.06 & 0.59 $\pm$ 0.15\\
TYC\,808-1232-1 & HERMES & 4910 $\pm$ 20 & 1.82 $\pm$ 0.06 & -0.91 $\pm$ 0.06 & 0.61 $\pm$ 0.07 & 0.5 $\pm$ 0.5 & 1.64 $\pm$ 0.08 & 1.13 $\pm$ 0.19\\
TYC\,5942-514-1 & HERMES & \referee{4323 $\pm$ 11} & 1.64 $\pm$ 0.05 & -0.43 $\pm$ 0.04 & 0.95 $\pm$ 0.09 & 1.04 $\pm$ 0.06 & 1.80 $\pm$ 0.05 & 1.21 $\pm$ 0.17\\
TYC\,6190-434-1 & HERMES & \referee{4076 $\pm$ 6} & 1.08 $\pm$ 0.03 & -0.61 $\pm$ 0.09 & 0.0 $\pm$ 0.3 & 0.35 $\pm$ 0.09 & 0.65 $\pm$ 0.06 & 0.38 $\pm$ 0.13\\
TYC\,8367-1842-1 & UVES & \referee{4675 $\pm$ 13} & 2.07 $\pm$ 0.03 & -0.44 $\pm$ 0.05 & 0.79 $\pm$ 0.07 & 0.67 $\pm$ 0.06 & 1.42 $\pm$ 0.03 & 0.97 $\pm$ 0.17\\
TYC\,8733-988-1 & UVES & \referee{4556 $\pm$ 11} & 1.94 $\pm$ 0.03 & -0.21 $\pm$ 0.04 & 0.59 $\pm$ 0.10 & 0.58 $\pm$ 0.05 & 1.28 $\pm$ 0.03 & 0.77 $\pm$ 0.13\\
\bottomrule
\end{tabular}
\tablefoot{
 The table only includes the abundances for the four s-process elements that are used in Sect.~\ref{sec:comparison} for the comparison with GALAH.
}
\end{small}
\end{table*}

\section{Spectrum analysis of the HR sample}\label{sec:specanalysis}

\subsection{Pre-processing}\label{sec:preprocessing}
After the spectra were reduced with the corresponding pipelines, we performed a series of steps to clean and prepare the data for a homogeneous spectroscopic analysis. For the UVES targets, we had a single spectrum per star, which was divided into two files: one file for the blue arm of the spectrograph, and the other file for the red arm. These spectra were first cleaned of cosmic rays and telluric absorption lines. The two spectral regions for each target were then merged into a single file, preserving the gap between them by setting the flux to zero in that region. The merged spectra were subsequently normalised using cubic splines computed over consecutive segments, and finally, the resulting spectra were corrected for the radial velocity. All pre-processing steps were performed with the spectral analysis tool \texttt{iSpec} \citep{iSpec,Blanco-Cuaresma19}, using its native functions along with the available telluric line lists and spectral masks.

For five of the HERMES targets, multiple exposures were acquired within the same observing run (in the same night or in two consecutive nights). Many more had multiple spectra covering a time span of years (see Sect. \ref{sec:binarity}), but only spectra obtained within the same night or in two successive nights were combined. This was done to avoid introducing spurious effects when we combined data of variable stars across different seasons. The individual spectra were cleaned of cosmic rays and telluric absorptions as described above and were then co-added as implemented in \texttt{iSpec}. A reference spectrum was chosen, and all other spectra of the same star were cross-correlated against it. \Ana{Finally, after the aligned spectra were continuum-normalised and combined through a weighted co-addition of their fluxes based on their S/N. This approach ensures that the final spectrum preserves the signal while reducing random noise contributions from individual exposures. Finally, the spectra of the HERMES targets with single exposures were also normalised and corrected for radial velocity with \ispec.}
Table \ref{table:HRtargets} lists the number of spectra we combined in each case and the final S/N. 

\subsection{Determining the atmospheric parameters}\label{sec:params}

The atmospheric parameters of our targets were also determined with \ispec\ \citep{iSpec, Blanco-Cuaresma19}, which generates synthetic spectra on the fly using several radiative transfer codes. For this project, we used the radiative transfer code Turbospectrum \citep{1998Alvarez,2012Plez}. The calculations assumed local thermodynamic equilibrium and employed 1D spherical MARCS model atmospheres \citep{2008Gustafsson}. Following the same strategy as in \cite{Blanco-Cuaresma19, 2020Casamiquela, VitaliEscorza24}, and \cite{Vitali24}, we derived stellar atmospheric parameters by fitting synthetic spectra. Our free parameters were \Teff, \logg, [M/H], \alphaM \Ana{, and line-broadening parameters. The latter were modelled using three quantities: the microturbulence velocity (\vmic), the macroturbulence velocity (\vmac), and the parameter $R$. $R$ does not correspond to the nominal instrumental resolution; rather, it is a broadening parameter used by the code to account for the overall line broadening, including contributions from rotation (\vsini) and macroturbulence (\vmac). } It is highly degenerate to separate these effect in \ispec\ \citep{iSpec}, and we therefore adopted a pragmatic approach \citep[similar to what is done in][]{Blanco-Cuaresma19} and fixed \vsini to 2 \kms, typical for giant stars, which are the bulk of our sample. \Ana{Then, we estimated \vmac from the empirical relation calibrated by the Gaia
ESO Survey working groups and provided in \citep{2014A&A...566A..98B}. In this way, $R$ accounted for the instrumental resolution and any differences between nominal and effective line broadening, ensuring that the results were not dominated by the assumed low \vsini\ value.} 

\Ana{As in GALAH DR4, the line list we used to determine the parameters was based on the Gaia-ESO line list \citep{Heiter2021}, which includes hyperfine splitting, with particular care taken to exclude weak lines or those affected by blends, telluric contamination, or uncertainties in continuum placement. Additionally, we adopted the solar abundances from \citet{Grevesse2007}.  Internal errors on the derived atmospheric parameters were computed from the covariance matrix of the fit, which accounts for the line‑to‑line scatter and the model sensitivity \citep[see][for details]{iSpec}. The final stellar parameters are presented in Table~\ref{table:HRresults}.}

\begin{figure*}[t]
    \centering   \includegraphics[width=0.9\linewidth]{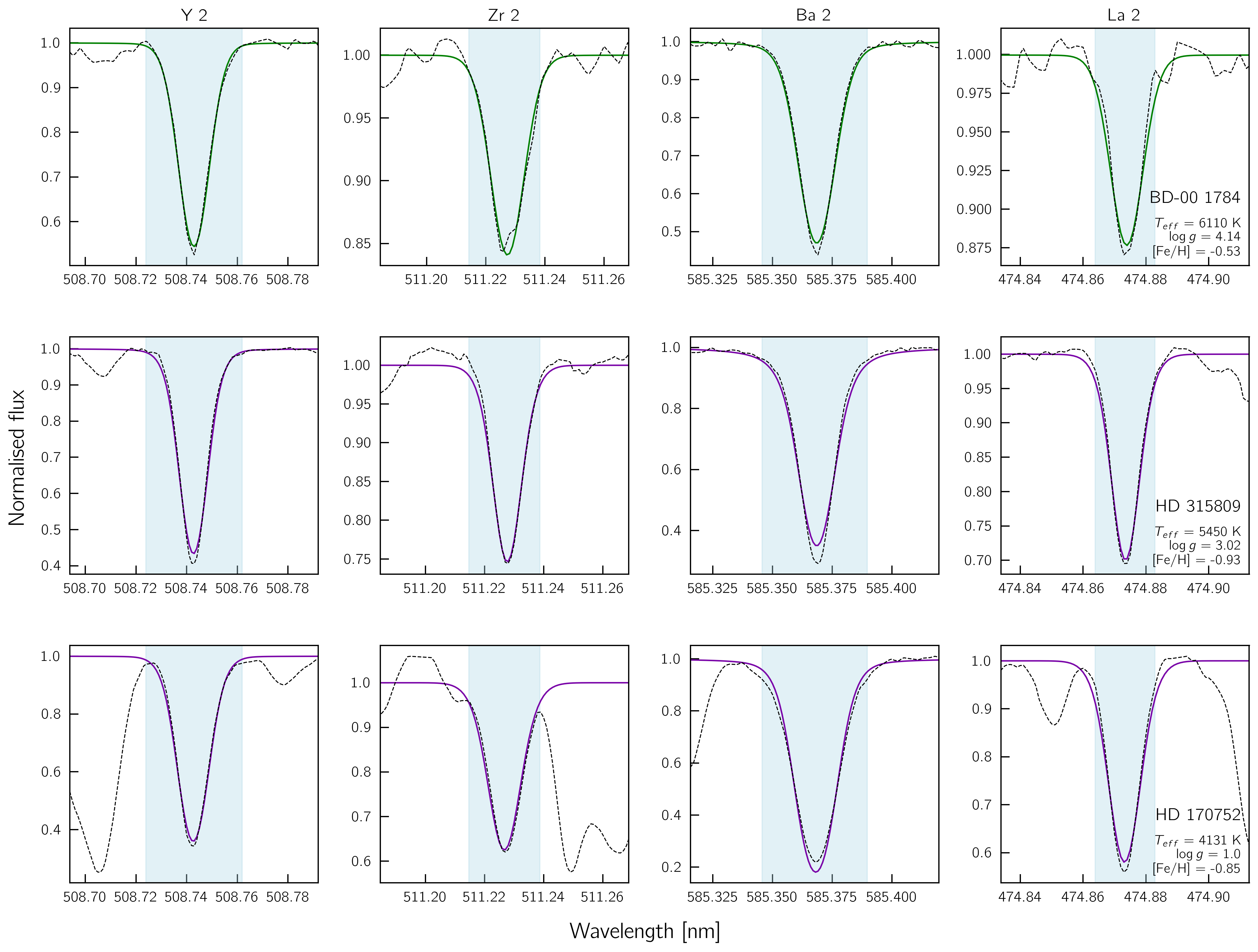}
    \caption{Example spectra of three stars from the HERMES (first row) and UVES (second and third rows) samples spanning different stellar parameters, illustrating the overall quality of our line fits. The observed spectrum is shown as a dashed black line, and the best-fitting synthetic spectrum \Ana{of the single absorption line} is overplotted with solid green (for HERMES) or purple lines (for UVES). The shaded region marks the wavelength interval we used to derive the abundances. From left to right, the panels display the \ion{Y}{II} line at 508.74 nm, the \ion{Zr}{II} line at 511.22 nm, the \ion{Ba}{II} line at 585.36 nm, and the \ion{La}{II} line at 474.87 nm. The names of the example stars and their corresponding atmospheric parameters are reported in each panel.}\label{fig:lines}
\end{figure*}

\subsection{Determining the chemical abundances}\label{sec:abundances}

To derive chemical abundances, we followed the same method as in \cite{VitaliEscorza24} and \cite{Vitali24}. First, the atmospheric parameters were fixed to the values determined in Sect.~\ref{sec:params}. We then performed spectral synthesis using the same radiative transfer code, atmospheric models, and line lists as employed for the stellar parameter determination. All our calculations were made in LTE. \Ana{However, GALAH DR4 uses 1D NLTE synthesis for some atomic lines \citep[BALDER code,][]{Amarsi18}. Of the four elements we chose to work with, they only used NLTE synthesis for Ba, for which they used the model atom from \citealt{Gallagher20} over the MARCS model atmosphere grid. To account for this, we applied the NLTE corrections from \cite{Korotin15} to our Ba line abundances as described below.}

\Ana{Even though we determined the individual abundances of other elements, with particular care in regions with blended lines, we only report the abundances of the four elements here that we used for this work.} The specific line selection for the heavy metals was constructed following the approach of \cite{VitaliEscorza24} and \cite{Karinkuzhi18} for barium and other s-process-rich stars. Since our ultimate goal was to compare our abundances with GALAH DR4, we also examined their line selection \citep{2025Buder} and included some of the lines they used. Below, we list the specific lines we used for the four s-process elements on which this work is based.

\begin{itemize}
    \item \textbf{Y:} We computed our abundances using the \ion{Y}{II} lines at 488.3682~nm, 490.0119~nm, 508.742~nm, 512.3211~nm, 520.0406~nm, 528.9815~nm, 540.2774~nm, 554.4611~nm, and 572.88865~nm. 
    \item \textbf{Zr:} We used the \ion{Zr}{I} lines at 480.587~nm, 481.563~nm, 482.804~nm, 612.744~nm, 613.455~nm, and 614.32~nm, and the \ion{Zr}{II} lines at 511.227~nm and 535.035~nm.
    \item \textbf{Ba:} We used the \ion{Ba}{II} lines at 585.36~nm and 649.70~nm, the same as are used by GALAH. Then, since GALAH DR4 uses NLTE synthesis for Ba, we applied the NLTE corrections provided for these two lines by \citet{Korotin15} after computing our Ba abundance per line. We used the stellar parameters we computed to determine the appropriate correction per line for each star. Finally, we computed the average abundance from these lines as for the other elements.
    \item \textbf{La:} We used \ion{La}{II} lines at 474.8726 nm, 480.4031 nm, 492.0965 nm, 529.082 nm, 530.3514 nm, and 639.0457 nm.
\end{itemize}

The mean abundance ratio $\left\langle \mathrm{[X/Fe]} \right\rangle$ for each element was computed using the line-by-line absolute abundances of each element. Most s-process elements have very few available and reliable lines, so each line was visually inspected to ensure the quality of the fit, and individual lines were removed on a star-by-star basis when our model did not reproduce them appropriately. We express our abundance values relative to the Sun, adopting the solar reference values of \citet{Grevesse2007}, as done for GALAH DR3 and DR4 \citep{2021Buder, 2025Buder}. GALAH uses zero-point corrections on their solar reference to match the literature (see Sect. 6.2.1. and Table 8 in \citealt{2025Buder}), but we decided to use the values published by \citet{Grevesse2007} and discuss the effect of these zero-point corrections in Sect.~\ref{sec:comparison-abu}. \Ana{The final abundances for Y, Zr, Ba, and La are presented in Table~\ref{table:HRresults}. Additionally, Fig. \ref{fig:lines} shows one line per element of interest for three example stars to illustrate the quality of our fits. \Teff, \logg, and \FeH of these targets are different enough to be representative of the sample.}

Abundance uncertainties in \texttt{iSpec} were estimated using two consistent methods \citep{iSpec}. When multiple spectral lines were available, the uncertainty was taken as the standard error of the line-by-line measurements. For elements with only a few lines, as in the case of our elements of interest, the uncertainties were estimated from the covariance matrix of the non-linear least-squares fit, which accounts for the sensitivity of the synthetic spectrum to parameter variations and the residuals between observed and synthetic spectra. \Ana{The uncertainties we report do not account for systematic errors, that is, the propagation of uncertainties on the stellar parameters. However, since our $\log g$ values and GALAH DR4 $\log g$ values are significantly different, Appendix~\ref{appendix:uncert} discusses the effect of varying $\log g$ on the measured abundances.}

\section{Comparison with GALAH DR4}\label{sec:comparison}
\subsection{Atmospheric parameters}\label{sec:comparison-params}

\begin{figure*}[t]
    \centering
    \includegraphics[width=\textwidth]{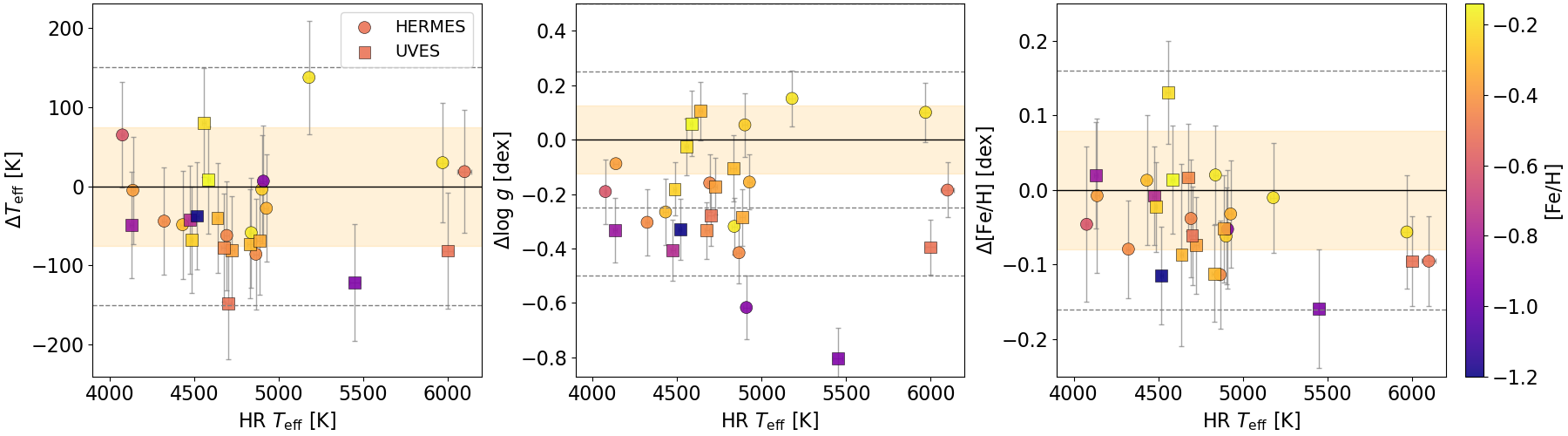}
    \caption{Difference (this work - GALAH DR4) between our HR results and the values published in GALAH DR4 \citep{2025Buder}. The panels show, from left to right, the differences in effective temperature, surface gravity, and metallicity. Each panel illustrates HERMES targets with circles and UVES targets with squares, and the symbols are colour-coded as a function of the derived metallicity. The yellow regions indicate where the difference would be within the average error bar ($1\sigma$), \Ana{calculated as the square root of the sum of the two error bars involved.} The dashed grey lines limit the region where the differences would be within $\pm2\sigma$ and $\pm3\sigma$ (only visible in the \logg\ plot since the negative offset there is considerable).}
    \label{Fig:delta_params}
\end{figure*}

Figure~\ref{Fig:delta_params} shows the difference between our \Ana{derived} spectroscopic parameters (Table~\ref{table:HRresults}) and the values published for the same stars in GALAH DR4 \citep{2025Buder} as a function of temperature. The differences were calculated as follows:
\begin{equation}
    \Delta P = P_{\rm this~work} - P_{\rm GALAH~DR4}, 
\end{equation}
where $P$ is, from left to right, \Teff, \logg, and [Fe/H].

The left panel shows that almost all $\Delta$\Teff\ values are within the $1\sigma$ region, indicating no significant differences between our temperatures and the catalogue values. The small vertical offset indicates that our temperatures are systematically lower than reported in GALAH DR4. The mean difference is about 30~K. A similar offset of 21~K was reported by \cite{2025Buder} when they compared their GALAH DR4 temperatures with the \textit{Gaia} FGK benchmark stars \citep{Jofre14, Jofre15, Jofre18, Heiter15}, with GALAH again reporting slightly hotter temperatures. Given the different methods and data, the differences are negligible.

\logg, however, shows larger differences. Again, our \logg\ values are systematically lower than the GALAH DR4 values, with the average difference being around $0.2$~dex. It is worth noting, however, that the method used to determine \logg\ in GALAH DR4 is very different to ours, since they applied post-processing corrections including non-spectroscopic information such as \textit{Gaia} EDR3 distances \citep{Bailer-Jones21} and photometry (see Sect.~5 in \citealt{2025Buder} for further details). When we compare our values to their purely spectroscopic \logg\ values, the scatter is still large, and some $\Delta$\logg\ values reach $\pm0.5$~dex, but the systematic offset is significantly reduced. \Ana{In Appendix~\ref{appendix:uncert}, we discuss the effect of varying \logg\ to better match GALAH DR4 on the abundances we derived.}

\begin{figure*}[t]
    \centering
    \includegraphics[width=0.8\textwidth]{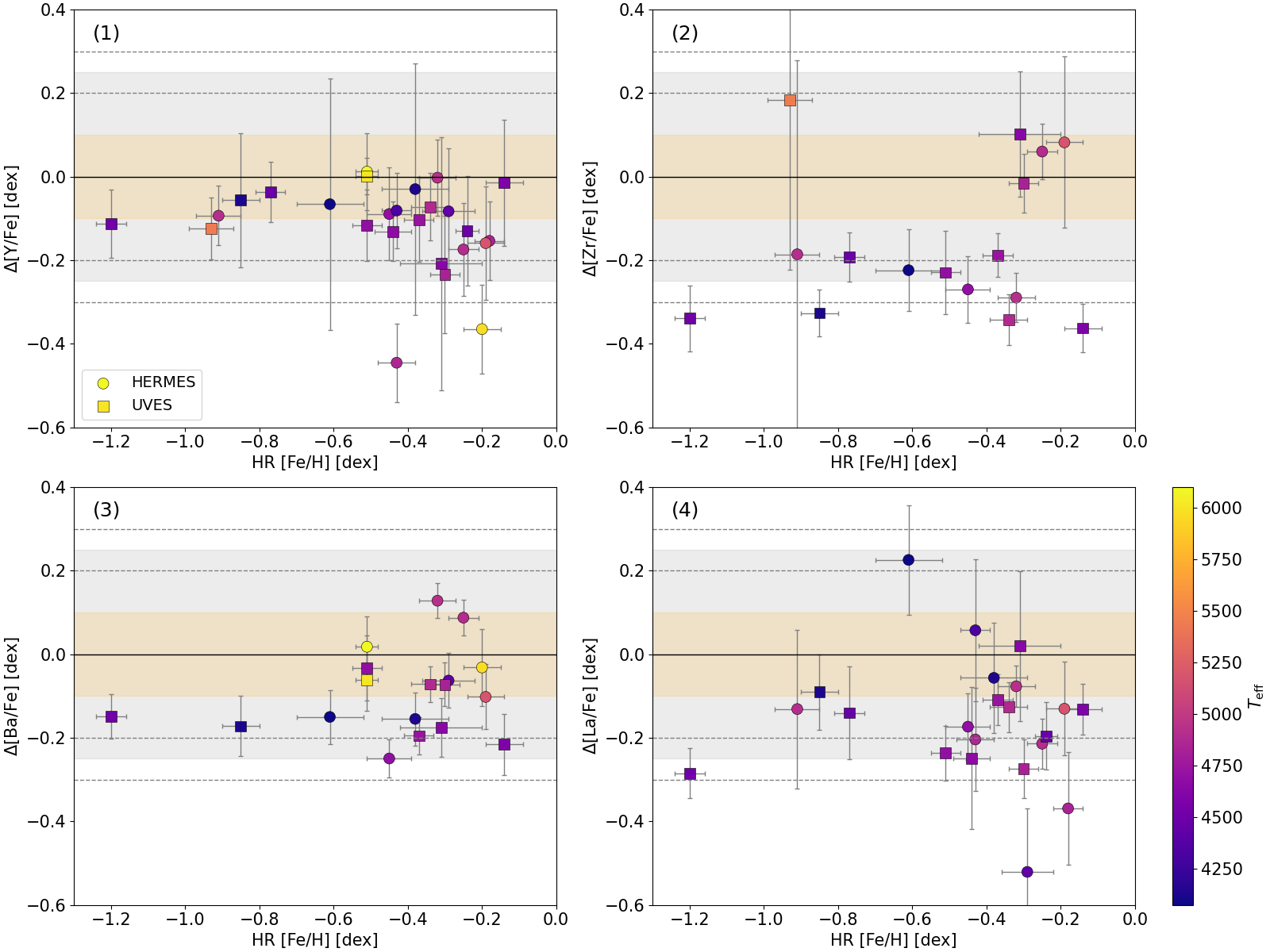}
    \caption{Same as Fig.~\ref{Fig:delta_params}, but for individual abundances. The panels show, from (1) to (4), differences in [Y/Fe], [Zr/Fe], [Ba/Fe], and [La/Fe]. This time, the symbols are colour-coded as a function of \Teff. The yellow regions indicate where the difference would be within $1\sigma$, and the dashed grey lines limit the $2\sigma$ and $3\sigma$ regions. We also show the $\pm0.25$~dex limit in grey that has been traditionally used to identify Ba and related stars.}
    \label{Fig:delta_abus}
\end{figure*}

Finally, the rightmost panel shows the metallicity differences between this work and GALAH DR4. As is the case for the temperature, the agreement is excellent, with all targets within a difference of $\pm0.15$~dex and most of them within the $1\sigma$ region of $\pm0.08$~dex. A negative systematic offset is again visible, but it is only 0.04~dex. An interesting metallicity effect to discuss is that the star that appears as an outlier in the three plots is HD\,315809 (observed with UVES), and it is among the most metal-poor stars in our spectroscopic sample ([Fe/H] = -0.93 $\pm$ 0.06~dex). TYC\,808-1232-1 (observed with HERMES), which also stands out in the $\Delta$\logg\ plot even though its $\Delta$\Teff\ and $\Delta$\FeH\ values are small, is also metal poor ($\rm{[Fe/H]}=-0.91\pm0.06$~dex). \Ana{\citet{2025Buder} stated that GALAH DR4 might perform more poorly at low metallicity. However, since it is rare to find large benchmark samples of metal-poor stars, they were unable to properly evaluate the effect. Our current sample is not statistically significant in this regime either, but building a new HR sample to evaluate the metal-poor regime in GALAH DR4 might be an interesting follow-up plan.}

\subsection{Chemical abundances}\label{sec:comparison-abu}

Figure~\ref{Fig:delta_abus} shows the differences between our abundances for Y, Zr, Ba, and La and those reported by GALAH DR4. Because our spectroscopic sample selection was made with GALAH DR3, not all stars have available abundances in GALAH DR4 for the four elements. This is simply due to the different method and the different quality cuts applied to the two \Ana{data releases}.

Even \Ana{though the agreement is good, there are again systematic offsets} between our results and the GALAH DR4 measurements. The average differences are $-0.12$, $-0.16$, $-0.09$, and $-0.16$ for [Y/Fe], [Zr/Fe], [Ba/Fe], and [La/Fe], respectively. Offsets of similar amplitudes are common between surveys and are generally caused by method systematics. For example, \cite{2023Heged} found similar offsets when they compared red giants in \Ana{\textit{Gaia}-ESO} with GALAH DR2 (i.e. $\Delta \rm{[Ba/Fe]} = 0.138$ and $\Delta \rm{[La/Fe]} = -0.140$). In our case, our differences with GALAH DR4 might also arise from the different lines used, since we used more lines for three out of the four elements. For example, the line choice for Zr strongly depends on spectral type and metallicity, requiring a careful line selection on a star-by-star basis that is not possible for large and automated surveys \citep[e.g.][]{Heiter2021,2022Kolomiecas,2023Sandford}. Moreover, the smallest offset is found for Ba, for which we used the same two lines. 
\Ana{We considered additional lines because our goal was not to reproduce GALAH DR4 results, but to identify these offsets between the survey and a star-by-star method based on HR spectra and to take them into account in the search for s-process-rich stars.}

\begin{figure*}[t]
    \centering
    \includegraphics[width=\textwidth]{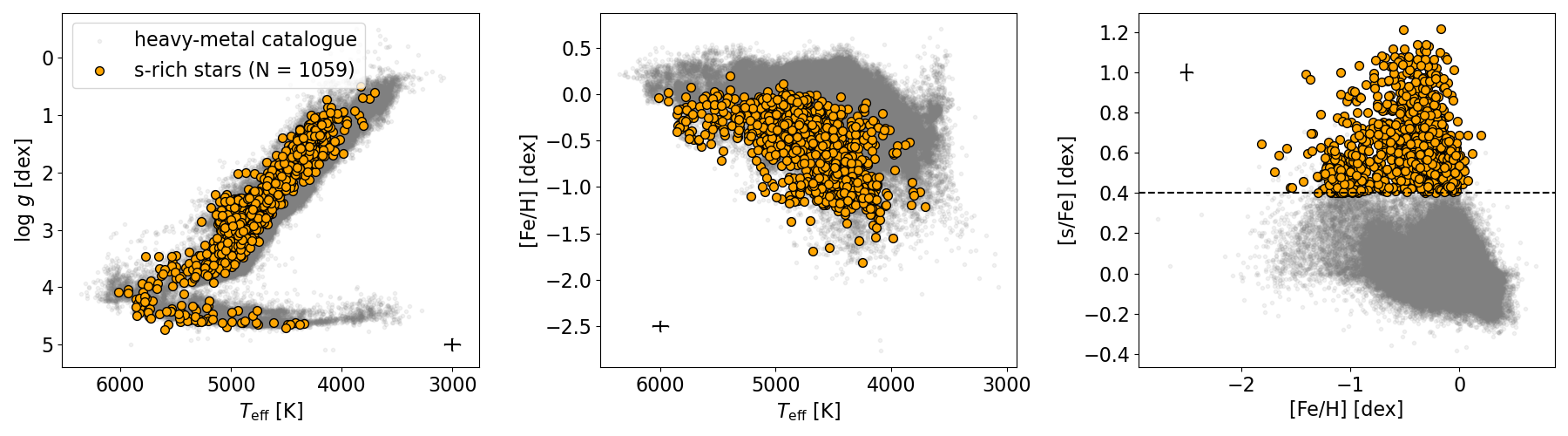}
    \caption{Kiel diagram, \Teff-[Fe/H] diagram, and [Fe/H]-[s/Fe] diagram for \Ana{our s-process-rich stars (orange circles). We plot it over the full “heavy-metal catalogue” (grey dots) for reference. Each panel includes the typical error bar that GALAH DR4 reports for these parameters.}}
    \label{fig:s-rich}
\end{figure*}

The abundances in Fig.~\ref{Fig:delta_abus} are plotted as a function of metallicity and colour-coded as a function of temperature so that trends or regions in the stellar parameter space become clear where the GALAH DR4 abundances match our values poorly. However, we do not identify significant biases within the parameter space covered by our HR sample. We also investigated possible trends with respect to \logg and $\Delta$\logg\ to evaluate the effect that our \logg discrepancy with GALAH DR4 might have on the abundance differences, but we again failed to find correlations.
 
\Ana{Finally, we note that the GALAH DR4 abundances were calculated using modified solar references (see Table 8 in \citealt{2025Buder}) to match several literature sources better. If we had use these modified solar references in our abundance calculations, we would have reduced the offset in [Y/Fe] and [Ba/Fe], while [Zr/Fe] would not change and the offset in [La/Fe] would slightly increase. However, these modified solar reference values are an empirical calibration for the GALAH survey and not a true solar reference, and they would artificially bring our abundances closer to their results. This would alter the best-matching abundance values we measured. We therefore decided not to use them.}

\section{S-process-rich stars in GALAH DR4}\label{sec:bastars}

\begin{figure*}[t]
    \centering
    \includegraphics[width=0.75\textwidth]{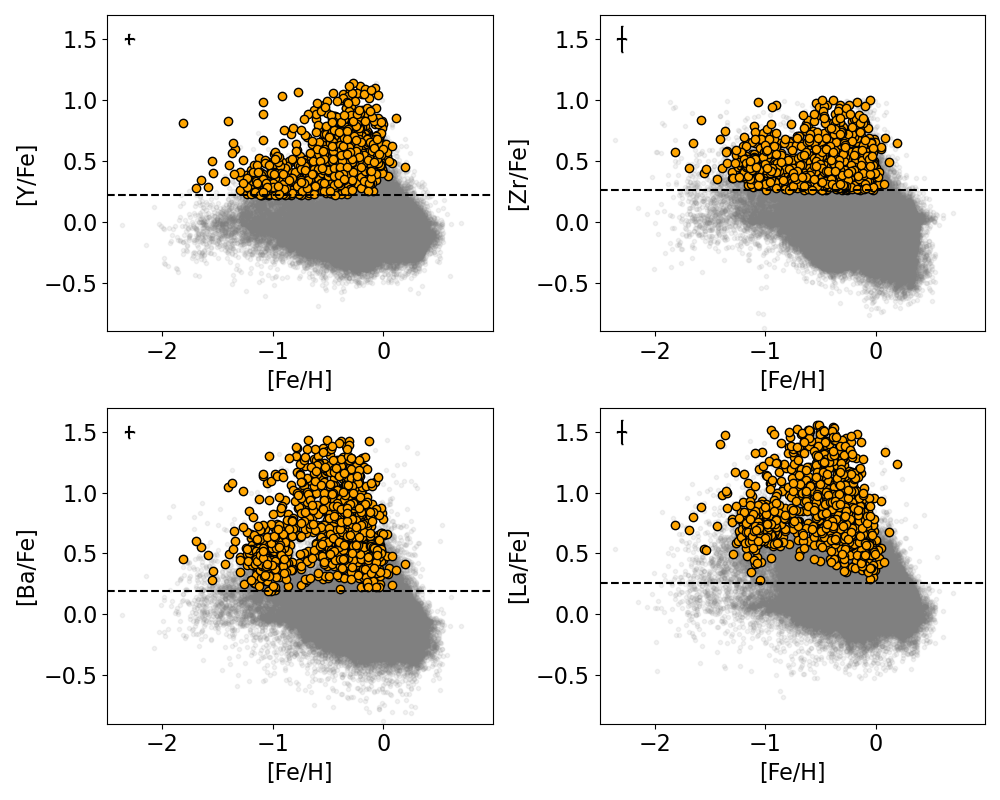}
    \caption{[Y/Fe], [Zr/Fe], [Ba/Fe], and [La/Fe] for our “s-rich catalogue”  (orange circles). As in Fig.~\ref{fig:s-rich}, \Ana{we plot it over the full “heavy-metal catalogue” (grey background dots) for reference. Each panel also includes the typical error bar that GALAH DR4 reports for these parameters and the selection thresholds listed in Sect~\ref{sec:bastars}.}}
    \label{fig:srich-abund}
\end{figure*}

\Ana{With the abundance offsets described above ($\Delta\rm{[Y/Fe]} = -0.12$, $\Delta\rm{[Zr/Fe]} = -0.16$, $\Delta\rm{[Ba/Fe]} = -0.09$, and $\Delta\rm{[La/Fe]} = -0.16$), }we then established specific thresholds that can help us identify s-process-enhanced stars in the GALAH DR4 catalogue with confidence. First, we defined [s/Fe] as the average abundance of the four s-process elements we selected as a base of our selection,
\begin{equation}
    \rm{[s/Fe]} = \frac{1}{4} \left( \rm{[Y/Fe] + [Zr/Fe] +[Ba/Fe] + [La/Fe]} \right).
\end{equation}

Traditionally, before the era of large spectroscopic surveys revolutionised the literature, Ba and related stars were not searched for \Ana{using thresholds on one element alone}. Normally, a threshold was set on [s/Fe] for this identification. For example, \cite{deCastro16} determined the abundances for the largest sample of confirmed Ba stars to date, defining a Ba star as a star with [s/Fe]~$>+0.25$. Their [s/Fe] was computed as the average of the individual abundances of Y, Zr, La, Ce, and Nd. \Ana{Additionally, \cite{Jorissen98} and \cite{Jorissen19} used [s/Fe]~$>+0.2$, considering two (Y and Nd) and four (Y, Zr, La, and Ce) elements, respectively.} They also established the threshold between mild and strong Ba stars at [s/Fe]~$>+1.0$. Following this, but considering the systematic offsets we detected, we defined an s-process-rich star as a star with [s/Fe]~$>+0.40$. Additionally, to ensure that the four measured elements had a super-solar abundance, we placed individual thresholds on each of them \Ana{based on the systematic offsets we measured (listed above) and the average error bar calculated when combining our errors with those reported by GALAH DR4 ($0.1$~dex)}. The summary of our constraints is the following:

\begin{figure}[t]
    \centering
    \includegraphics[width=0.49\textwidth]{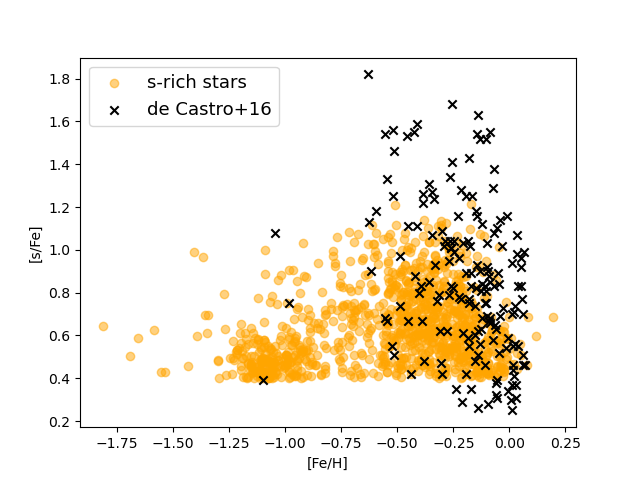}
    \caption{Comparison between our “s-rich catalogue”  (orange circles) and the Ba stars presented in \cite{deCastro16} (black crosses) in the [s/Fe]-metallicity plane.}
    \label{fig:comparison-sFe}
\end{figure}

\begin{itemize}
    \item $\rm{[s/Fe]} > 0.40$
    \item $\rm{[Y/Fe]} > 0.22$
    \item $\rm{[Zr/Fe]} > 0.26$
    \item $\rm{[Ba/Fe]} > 0.19$
    \item $\rm{[La/Fe]} > 0.26$.
\end{itemize}

Applying these thresholds to the “heavy-metal catalogue” (see Sect.~\ref{sec:GALAHprep}), we determined a new subset of GALAH DR4 targets that we refer to as the “s-rich catalogue” \footnote{The final catalogue is available in electronic form at the CDS via \url{http://cdsweb.u-strasbg.fr/cgi-bin/qcat?J/A+A/}}. It contains \referee{1059 stars} that make up our sample of s-process-polluted candidates. This sample represents \referee{0.664\%} of the “heavy-metal catalogue”, \Ana{which is slightly below the 1\% traditionally mentioned for Ba stars  \citep{MacConnell72, NorthDuquennoy91}. However, this is consistent with the fact that our thresholds for defining a polluted candidate are more conservative than the $\rm{[s/Fe]} > 0.20 - 0.25$~dex traditionally used. }

\Ana{If we were to relax our thresholds, use [s/Fe]~>~0.25, and remove the individual thresholds set on each element, we would find \referee{more than 5300 candidates, representing 3.3\% of the “heavy-metal catalogue”}. However, as clearly stated above, we did not aim for a complete but for a pure sample, and relaxing the thresholds would lead to many impostors in the final sample. Finally, with our original thresholds, when we focus on the metal-poor area of our metalicity range, that is, below the Ba-star regime, \referee{we find that $\sim$3\% of the “heavy-metal catalogue” are s rich.}} The full GALAH DR4 table for this subset of stars will be made available as supplementary material.

Figure~\ref{fig:s-rich} shows this new sample in the Kiel diagram (\logg\ vs. \Teff; left), in a [Fe/H]-\Teff diagram, and in the [s/Fe]-[Fe/H] plane. Two interesting features deserve to be highlighted from these plots. The first feature is the lack of s-rich stars among cool stars (most of them have \Teff\ > 4000~K), which is a product of our abundance quality cuts. \Ana{We removed a few molecule-dominated targets from our HR sample, but we did not apply any temperature threshold to the GALAH data. As discussed by \cite{2025Buder}, even though GALAH DR4 can model some molecular features and infer C and N abundances from these, other atomic abundances are no longer reliable when molecules begin to dominate the spectra. This means that by imposing good-quality flags on four chemical elements, we reduced the chances to find molecule-dominated s-rich stars. The second feature is the lack of s-rich stars with super-solar metallicities. This is likely caused by the combinations of a physical effect and an observational bias. The nucleosynthesis s-process is less efficient inside AGB stars with super-solar metallicities \citep[e.g.][]{Karakas14}, and not many s-process-polluted stars have been characterised in this area of the parameter space. Additionally, at high metallicities, several s-elements are already enhanced due to the chemical evolution of the Galaxy, and many spectral lines are intrinsically stronger and even saturated, which makes an abundance determination, especially with automatic methods, more challenging.}

Finally, Fig.~\ref{fig:srich-abund} shows the individual abundances reported by GALAH DR4 for the four elements of interest, [Y/Fe], [Zr/Fe], [Ba/Fe], and [La/Fe]. We note that each individual panel contains stars with a high overabundance of that specific element, but that are not part of our final sample \Ana{ because some stars have high abundances in one or more s-process elements, but have subsolar abundances in the other(s), and the thresholds set on the individual abundances excluded them from consideration. Since individual s‑process abundances are expected to increase in concert in AGB mass‑transfer products, the presence of such stars underscores the risk of relying on a single element for a reliable identification of chemically peculiar stars.}

\Ana{We cross-matched our GALAH DR4 candidates with the list of 486 Ba star candidates recently published by \cite{Yang26}, who also used GALAH DR4. \referee{There are 166 stars in common between the two samples}, which is a consequence of the different choices made when flagging a GALAH star as an s-process-rich candidate. \cite{Yang26} only used two s-process elements in their selection method (Ba and La), and they used a less conservative threshold on [s/Fe] ($=0.25$~dex). However, they included thresholds on [Ba/Eu] and [La/Eu] to restrict the contribution of r-process (rapid neutron-capture process) pollution, which we decided not to adopt. Including a quality threshold on the abundance of Eu as an additional restriction would cut our candidate sample by a factor of $\sim$2, and the fact that we can reproduce the [hs/ls] distribution with s-process nucleosynthesis models  (see Sect. \ref{sec:comparison-ba} below) indicates that this is not necessary.}

\section{Discussion}\label{sec:discussion}
\subsection{Comparison with \Ana{confirmed} Ba star samples}\label{sec:comparison-ba}

\begin{figure}[t]
    \centering
    \includegraphics[width=0.49\textwidth]{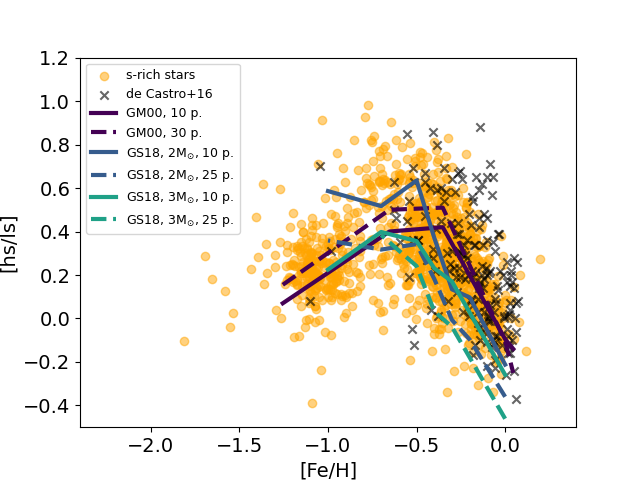}
    \caption{\Ana{Comparison between our “s-rich catalogue” (orange circles) and the Ba stars presented in \cite{deCastro16} (black crosses) in the [hs/ls]-metallicity plane. For comparison, we overplot in purple the AGB nucleosynthesis yields published by \citet[][GM00]{GorielyMowlavi00} after 10 (solid line) and 30 (dashed line) identical thermal pulses, and we show a set of more recent models from \citet[][GS18]{GorielySiess18} for 2~\Msun\ and 3~\Msun\ stars in blue and green, respectively. For the latter, we also show the yields after 10 (solid line) and 25 (dashed line) thermal pulses.}}
    \label{fig:comparison-hsls}
\end{figure}

Figures~\ref{fig:comparison-sFe} and \ref{fig:comparison-hsls} show our sample in two of the most classical planes for characterising Ba and related stars. In both figures, we overplot the sample from \cite{deCastro16}, since it is the largest and most systematically analysed sample of barium stars to date (see as well the expansions published by \citealt{Roriz21-Rb, Roriz21-more, Roriz24-W, Roriz25-CNO}). \Ana{Before we created these plots, we cross-matched our s-rich candidate sample with their sample, and there is only one star in common: HD\,122687 (Gaia DR3 6094094603419398912).}

In Fig.~\ref{fig:comparison-sFe} we plot [s/Fe] as a function of metallicity, and considerable differences with the de~Castro sample are clear. First, the discrepancy at low [s/Fe] is caused by the construction of our sample. To avoid polluting our sample with weak candidates, we set a threshold to select stars with $\rm{[s/Fe]} > 0.40$. However, \cite{deCastro16} had high-resolution spectra and analysed their sample star by star, so they were able to relax that threshold, which was $\rm{[s/Fe]} > 0.25$. \Ana{Then, our sample shows a lower density of stars with $-0.8<\rm{[Fe/H]<-0.6}$ and a second well-populated group with $-1.3<\rm{[Fe/H]<-0.8}$.} \cite{deCastro16} did not have many stars in this metallicity range because they built a catalogue of Ba stars, and typically, the Ba star classification is only used when $\rm{[Fe/H]} \gtrsim -0.5$. At lower metallicities, the equivalent s-process-polluted targets appear in the literature as CH or CEMP-s stars and were not part of their target selection, while we did not restrict the metallicity values. 

Finally, the top part of the y-axis in Fig.~\ref{fig:comparison-sFe} shows an interesting difference because our [s/Fe] values do not reach more than 1.2~dex. Our definition of [s/Fe] considers Y, Zr, Ba, and La, while theirs considered Y, Zr, La, Ce, and Nd, so some differences are expected. \Ana{Even though [s/Fe] is calculated as an average abundance and the element selection should not have such a strong effect, different s-process elements contribute more or less significantly to the r-process. A different distribution of the r-contribution to the elements in each list might partly explain a discrepancy with the literature [s/Fe] values. \referee{Additionally, the [La/Fe] abundances given in \cite{deCastro16} were updated by \citet{Roriz21-more}, who reported lower abundances on average. We recomputed [s/Fe] with the new La abundances to determine the effect on [s/Fe]. While some literature targets with high [s/Fe] show lower values, this is not the case for all, and the literature [s/Fe] values still reach higher values than were obtained for the GALAH s-rich sample. The figures only include the abundances reported by \cite{deCastro16} for simplicity.} Finally, another possible explanation is that} GALAH DR4 neural networks are not prepared for the extremes in stellar populations \citep{2025Buder}, and the quality cuts applied by them and by us excluded stars with the most extreme abundances.

Figure~\ref{fig:comparison-hsls} shows [hs/ls] as a function of metallicity, with [hs/ls] defined as
\begin{equation}
    \rm{[hs/ls]} = \rm{[hs/Fe] - [ls/Fe]},
\end{equation}
where [hs/Fe] is the average abundance of the heavy s-elements (Ba and La), and [ls/Fe] is the average abundance of the light s-elements (Y and Zr). Here, the differences with the de~Castro sample, in the [Fe/H] range from 0.0 to -0.5, are much smaller. While the exact trend that reaches higher [hs/ls] values for lower metallicity stars (within the Ba star regime) is slightly different, the range covered is the same and the differences can be explained \Ana{by the different methods and s-process elements included in the averages}. At lower metallicities, below -0.6, [hs/ls] clearly drops in our sample, clustering between [hs/ls] of around 0.0 and [hs/ls] of around 0.4, without showing any clear trend. This drop in [hs/ls] at lower metallicities has been predicted by some nucleosynthesis models. \Ana{In particular, we show in Fig.~\ref{fig:comparison-hsls} the AGB nucleosynthesis yields published by \cite{GorielyMowlavi00} for different surface enrichments, that is, after typically 10 to 30 identical thermal pulses. These models represent five stars with different metallicities and masses (between 1.5 and 3.0~\Msun) and consider the parametrised partial mixing of protons from the AGB envelope into the C-rich layers at the time of the third dredge-up. This partial mixing of protons leads to an s-process that is strongly sensitive to the stellar metallicity, and it in particular transforms material into lead at metallicities [Fe/H]$<-1$. This modelling is shown in Fig.~\ref{fig:comparison-hsls} to give rise to a bell-shaped [hs/ls] as a function of metallicity (we refer to \cite{GorielyMowlavi00} for the details of the models).}

\Ana{Additionally, we show in Fig.~\ref{fig:comparison-hsls} a more realistic prediction from \cite{GorielySiess18} for 2~\Msun\ and 3~\Msun\ AGB model stars. In this case, the partial mixing of protons at the time of the third dredge-up was modelled by a consistent diffusive overshoot prescription (see \citealt{GorielySiess18} for more details) within the STAREVOL stellar evolution code \citep{Siess00, Siess06, Siess08, SiessArnould08}. Globally, a similar behaviour of [hs/ls] as a function of the stellar metallicity is observed. Fig.~\ref{fig:comparison-hsls} confirms the key effect of the stellar metallicity on the s-process efficiency through the [hs/ls] proxy. This comparison between observations and s-process models in AGB stars confirms that the stellar metallicity plays a crucial role in determining the neutron-to-seed ratio available to the s-process nucleosynthesis, with primary $^{13}$C produced by the partial mixing of protons into the $^{12}$C-rich region at the time of the third dredge-up as the main neutron source.} \referee{Additionally, the fact that AGB models with 2 and 3~\Msun\ cover the full s-rich sample well strengthens the idea that low-mass (<~3~\Msun) AGB stars pollute Ba and related stars, as suggested by \citet{Lugaro03, Lugaro12, Lugaro16, Cseh18, Karinkuzhi18,Dimoff24}, among others.}

\subsection{Binarity}\label{sec:binarity}
Barium and related stars are binaries by formation, and, as discussed in Sect.~\ref{intro}, this was well established a few decades ago (see the recent reviews by \citealt{Escorza23SEA} and \citealt{M&P2025}). Then, an obvious next step for our sample would be to check for binarity. To do this, we queried our \referee{1059} candidates in the \textit{Gaia} DR3 Non-Single Star (NSS) catalogue \citep{GaiaDR3-NSS}. We found that the periods and eccentricities for only 77 of them ($\sim$7\%) are published by combining astrometric and spectroscopic solutions, while \referee{72} more are detected as potential binaries due to radial velocity trends or astrometric accelerations. Finally, \referee{fewer than 24\% (250 stars)} have a renormalised unit weight error (RUWE) value in \textit{Gaia} EDR3 \citep{GaiaEDR3} higher than 1.4, which is the value traditionally used to flag binary stars \citep[e.g.][]{Lindegren18, Belokurov20, Stassun21, Kervella22, Godoy-Rivera25}. These small numbers, however, are not completely surprising given that the periods of known s-process-polluted stars are up to several decades, far beyond the reach of the \textit{Gaia} mission, and many confirmed Ba or related stars with fully determined orbits have not been detected as binaries by \textit{Gaia} either \citep{Jorissen19EWASS, Escorza19EWASS, EscorzaDeRosa23}. \Ana{In any case, for the comparison, we calculated the same percentages for the \referee{159,464 stars} in the “heavy-metal catalogue”, and the binary fraction is significantly smaller. \referee{Fewer than 4\% of the stars have NSS~$\neq$0, and fewer than 10\% have RUWE~$\geq1.4$.}}

\begin{figure}[t]
    \centering
    \includegraphics[width=0.45\textwidth]{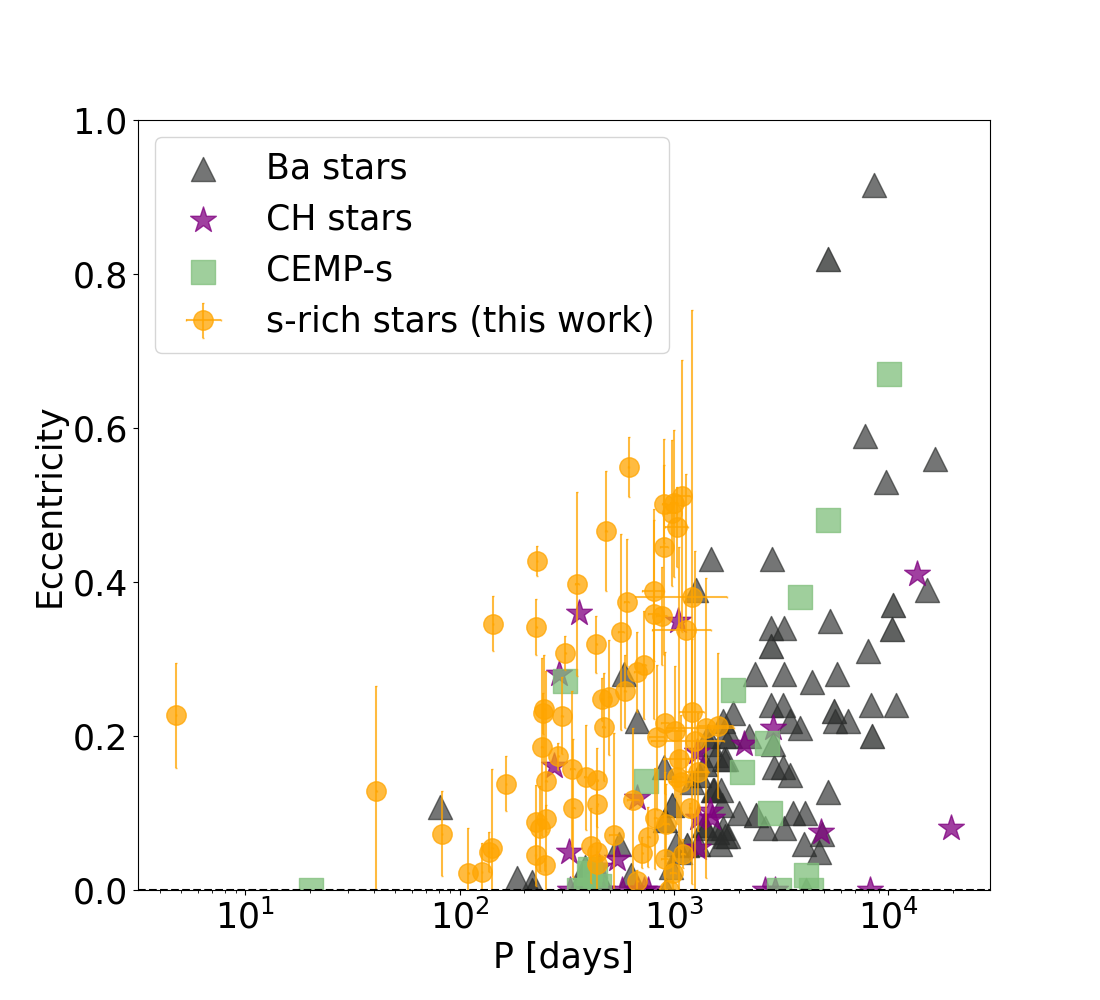}
    \caption{Eccentricity-period diagram of the 77 stars in our “s-rich catalogue” with an orbital solution in the \textit{Gaia} NSS catalogue \citep[orange circles][]{GaiaDR3-NSS}. We also include for comparison the Ba stars from \cite{Jorissen19, Escorza19, North00, EscorzaDeRosa23} as grey triangles, the CH stars from \cite{Jorissen19,Escorza19} as purple stars, and the CEMP-s from \cite{Jorissen16, Hansen16} as green squares.}
    \label{fig:eP}
\end{figure}

Figure~\ref{fig:eP} shows the 77 stars in our “s-rich catalogue” with periods and eccentricities in the NSS catalogue together with the Ba stars published by \cite{Jorissen19, Escorza19, North00, EscorzaDeRosa23}, the CH stars from \cite{Jorissen19,Escorza19}, and the CEMP-s from \cite{Jorissen16, Hansen16} for reference. \Ana{We cross-matched our s-rich candidates also with these samples, and there are no targets in common.} While our longest period is just below 1600~days, the other three samples cover a much wider range on the x-axis, \Ana{which is just a product of the limited baseline of the \textit{Gaia} mission, as mentioned above.} Additionally, the figure shows that within the range covered by all samples, the new \textit{Gaia} orbits seem to reach higher eccentricities than the known Ba, CH, and CEMP-s stars. \Ana{However, the error bars on the eccentricity, especially at long periods, are significant. }When \textit{Gaia} DR4 is published, we will expand the \textit{Gaia} eccentricity-period diagram of the “s-rich catalogue” and investigate this feature with the new data.

Additionally, there are a few sources with uncomfortably short periods, but Gaia DR3 2619209256264051712 is particularly problematic. \Ana{With $P=4.7628\pm0.0011$~days, it must have been subject to severe orbital shrinkage during the binary interaction if this object hosts a WD because an AGB star does not fit in the current orbit. Additionally, with $e=0.23\pm0.07$, common-envelope evolution after the mass-transfer episode is also unlikely for this system unless some eccentricity-pumping mechanism acted after the Roche-lobe overflow. The system might be a triple, with the WD companion in an outer orbit, as is the case for example for the Ba stars HD\,48656 \citep{Escorza19} and HD\,218356 \citep{EscorzaDeRosa23}.} The abundances reported by GALAH are the following: [Y/Fe]$=0.68\pm0.03$~dex, [Zr/Fe]$=0.52\pm0.06$~dex, [Ba/Fe]$=0.53\pm0.04$~dex, and [La/Fe]$=0.57\pm0.04$~dex \citep{2025Buder}. Hence, it passes all our thresholds for a mild Ba star candidate. High-resolution spectroscopy to confirm the abundances and high-cadence RV monitoring to confirm the orbit are necessary to explain the pollution and the orbital parameters of this target.

\begin{figure*}[t]
    \centering
    \includegraphics[width=0.24\textwidth]{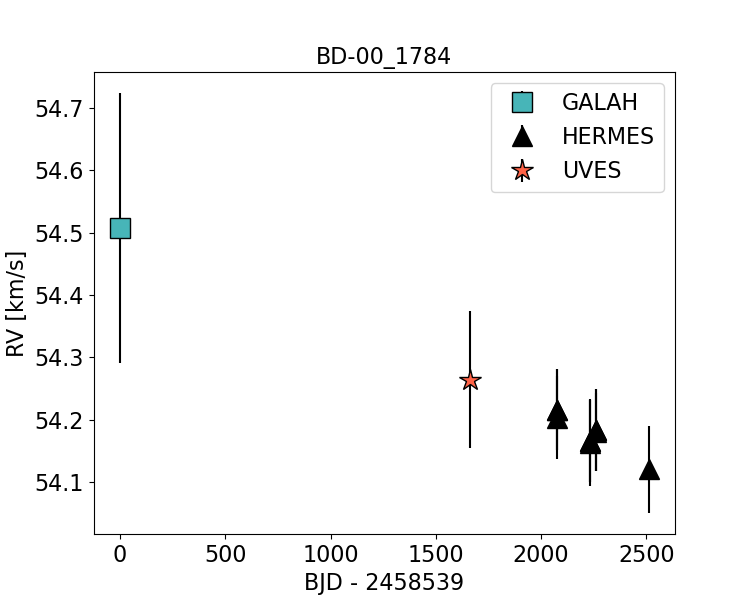}
    \includegraphics[width=0.24\textwidth]{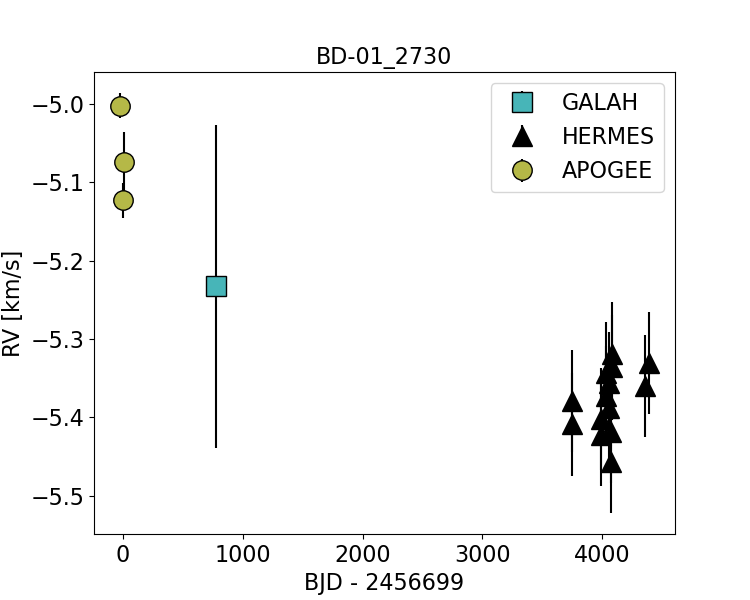}
    \includegraphics[width=0.24\textwidth]{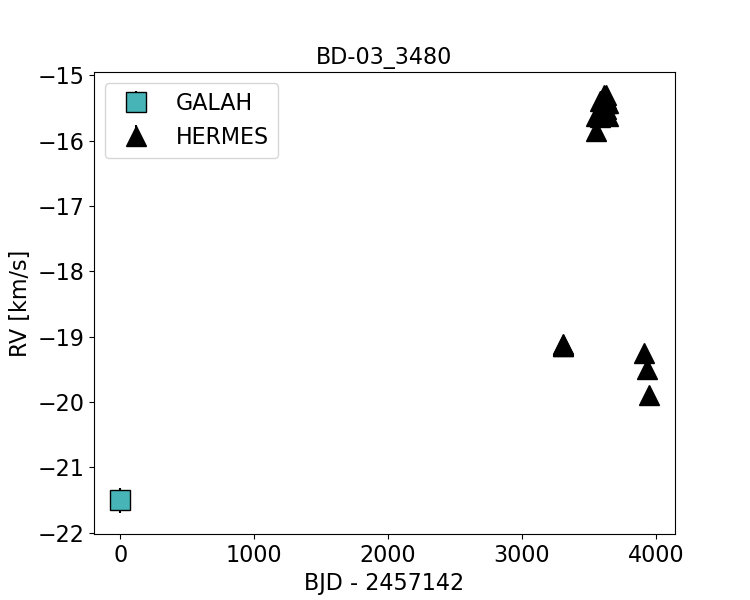}
    \includegraphics[width=0.24\textwidth]{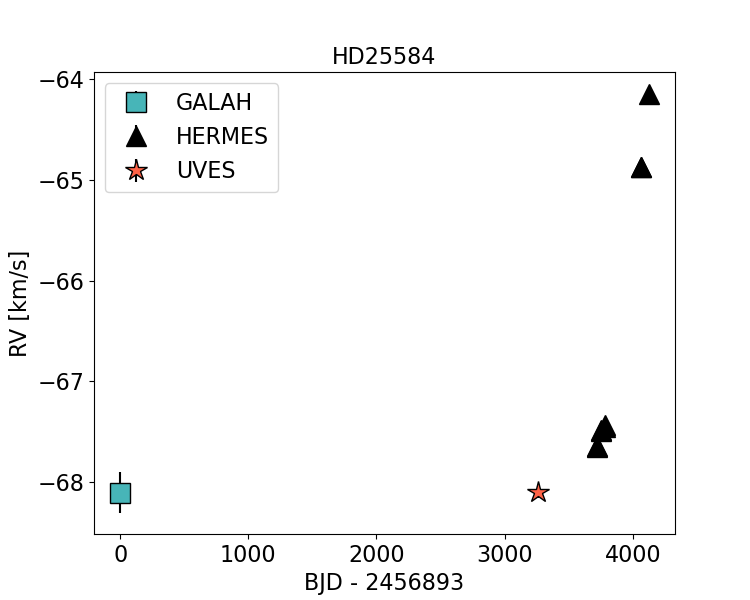}
    \includegraphics[width=0.24\textwidth]{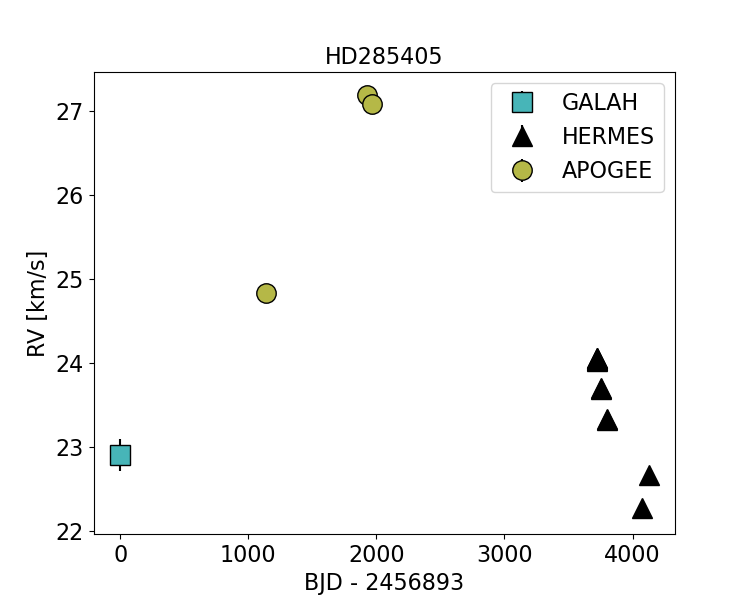}
    \includegraphics[width=0.24\textwidth]{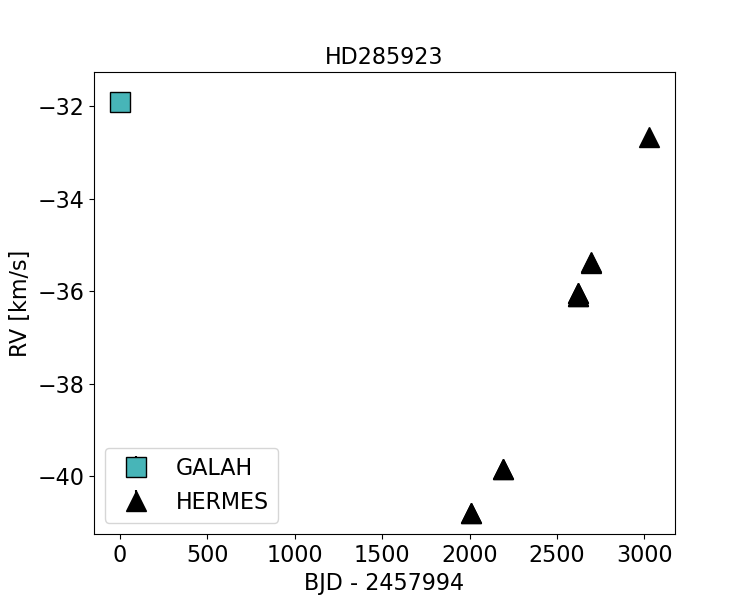}
    \includegraphics[width=0.24\textwidth]{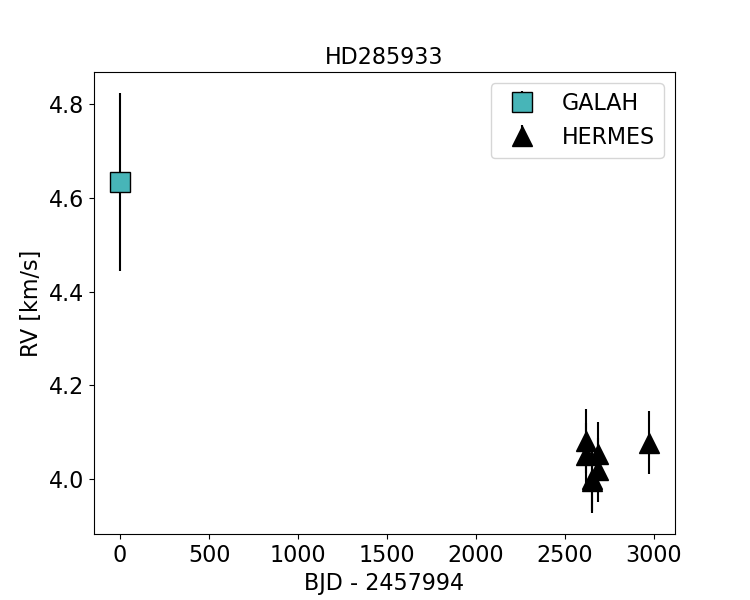}
    \includegraphics[width=0.24\textwidth]{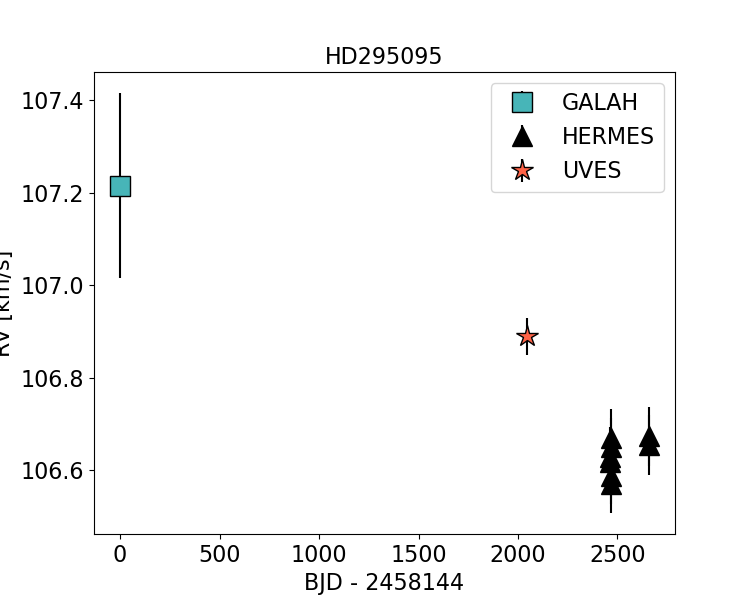}
    \includegraphics[width=0.24\textwidth]{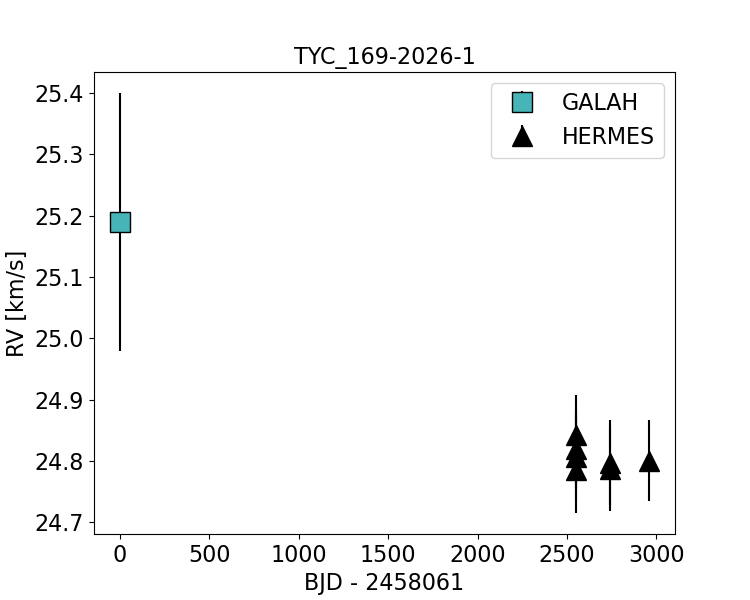}
    \includegraphics[width=0.24\textwidth]{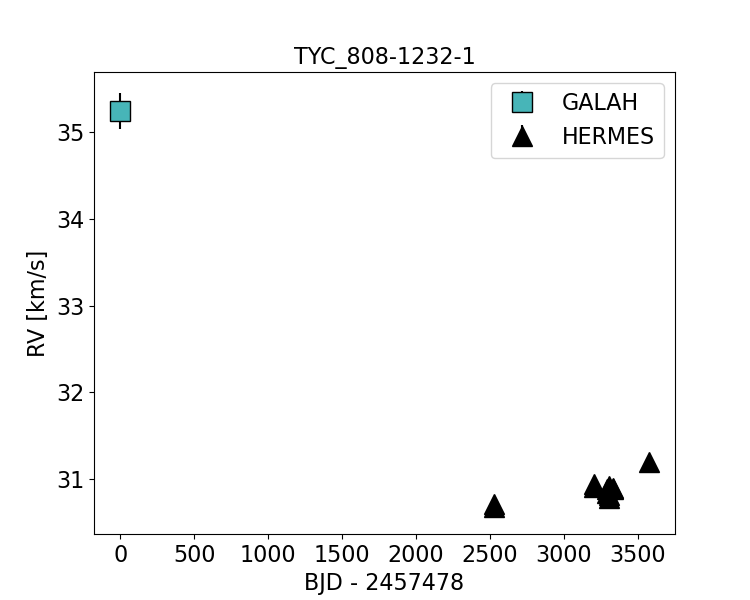}
    \includegraphics[width=0.24\textwidth]{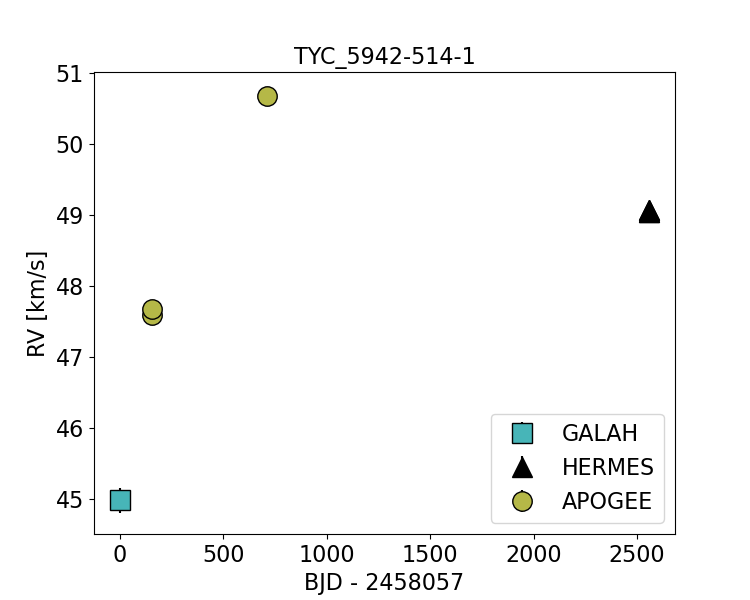}
    \includegraphics[width=0.24\textwidth]{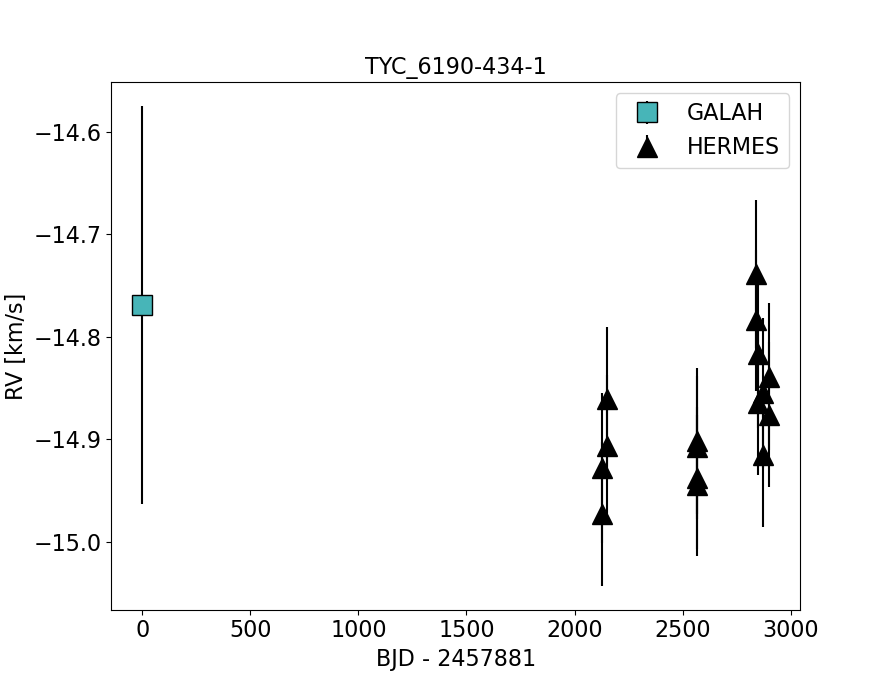}
    \caption{Radial velocity variability of the stars monitored with HERMES. TYC\,179-2074-1 is not included because we only have one spectrum.}
    \label{fig:RVsHERMES}
\end{figure*}

Finally, concerning binarity and the limitations of the \textit{Gaia} mission for long-period systems, we highlight that only three of our HERMES targets (HD\,25584, TYC\,5942-514-1, and TYC\,808-1232-1) are detected as binaries by \textit{Gaia}, and none of them has a published orbital solution.
However, we have a few spectra per target, and we illustrate in Fig.~\ref{fig:RVsHERMES} their radial velocity variability. \Ana{The HERMES baseline is still short, but when the GALAH RV value and the values derived by APOGEE \citep{APOGEE, Abdurrouf2022} or from our UVES spectra are included when possible, most of them show clear and significant trends that will result in newly characterised long-period binaries with enough telescope time.}

\section{Summary and main conclusions}\label{sec:summary}
We presented a sample with \referee{1059} GALAH DR4 stars that are very promising candidate s-process-polluted stars. This sample is almost five times larger than the number of currently confirmed polluted stars. \Ana{The metallicities of about 700 of our candidates are representative of Ba stars ($\rm{[Fe/H]}\gtrsim-0.5$), while most of the rest have $-1.25\lesssim\rm{[Fe/H]}\lesssim-0.5$, which would traditionally classify them as CH stars. The most important difference between our candidates and other similar works on this or other surveys is that we applied conservative thresholds that were informed by a critical validation of the GALAH DR4 abundance accuracies based on our analysis of high-resolution spectra.} We observed 24 stars from a subset of the GALAH catalogue that passed several quality checks with the UVES and the HERMES spectrographs. We used the public spectral software \ispec\ \citep{iSpec, Blanco-Cuaresma19} to determine their stellar parameters and individual abundances and compared our values for \Teff, \logg, \FeH, [Y/Fe], [Zr/Fe], [Ba/Fe], and [La/Fe] with those published by GALAH DR4 \citep{2025Buder}. This comparison helped us define the following thresholds on [s/Fe] and on the abundance of four s-process elements: $\rm{[s/Fe]} > 0.40$, $\rm{[Y/Fe]} > 0.22$, $\rm{[Zr/Fe]} > 0.26$, $\rm{[Ba/Fe]} > 0.19$, and $\rm{[La/Fe]} > 0.26$. We use them to select our \referee{1059 s-process-polluted candidates from the 159\,464 stars with which we we started. These candidates represent 0.66\% of the initial sample}, which is slightly below the 1\% traditionally mentioned for Ba stars  \citep{MacConnell72, NorthDuquennoy91}. \Ana{However, this is consistent with the fact that our thresholds for defining a polluted candidate are more conservative than the $\rm{[s/Fe]} > 0.20 - 0.25$~dex used traditionally. The occurrence rate increases at lower metallicities up to 3\%.}

We compared our polluted candidates with their confirmed counterparts \citep{deCastro16}, and while we found a few interesting differences in the [s/Fe]-[Fe/H] plane, their similarities in the [hs/ls]-[Fe/H] plane are remarkable. Our sample appears to contain a subset of Ba star candidates, with metallicities down to $\sim -0.6$~dex, which follow the same trend in the [hs/ls]-[Fe/H] plane as the Ba stars from \cite{deCastro16}. However, the sample also contains another subset of stars at lower metallicities that cluster at [hs/ls] from 0 to 0.4, which is below the most metal-poor Ba stars, which reach [hs/ls]$\sim0.9$. \Ana{This drop in the [hs/ls]-[Fe/H] plane at low metallicity was predicted by \cite{GorielyMowlavi00} and \cite{GorielySiess18}, among others.} Additionally, we queried the \textit{Gaia} DR3 NSS catalogue \cite{GaiaDR3-NSS} and discovered that only $\sim7\%$ of our \referee{1059} polluted star candidates have published orbital solutions, and fewer than 24\% have RUWE~$\geq1.4$. This is expected because the time baseline of the \textit{Gaia} mission and the long periods that these AGB interaction products can reach; however, it is important to highlight these limitations. \Ana{We also highlight that even though the binary fraction is low, it is higher than in the comparison “heavy-metal catalogue”, where fewer than 4\% of the stars have NSS~$\neq$0 and fewer than 10\% have RUWE~$\geq1.4$.}

\Ana{We cross-matched our candidates with the largest samples of confirmed Ba stars and found only one candidate in common. We also crossm-atched them with the candidates recently published by \cite{Yang26}, who also used GALAH DR4, and we have 166 stars in common. This means that our sample adds more than 900 new AGB-mass-transfer candidates.} More importantly, these \referee{1059} stars have been homogeneously analysed with a single method. The unique feature of GALAH, which published the abundances of eight s-process elements, opens windows for properly comparing nucleosynthesis models. Moreover, with future \textit{Gaia} releases, we will expand Fig.~\ref{fig:eP} towards longer periods and will be able to also test binary evolution models. Finally, what we learned from GALAH DR4 can be applied with caution to other large spectroscopic surveys such as APOGEE \citep{APOGEE} or \textit{Gaia} \citep{GaiaEDR3summary}, which provide abundances of a single s-process elements (Ce) for a much larger number of stars.\\

\noindent \textbf{Data availability:} The final “s-rich catalogue” is available in electronic form at the CDS via anonymous ftp to cdsarc.u-strasbg.fr (130.79.128.5) or via \url{http://cdsweb.u-strasbg.fr/cgi-bin/qcat?J/A+A/}
 
\begin{acknowledgements}
AE and SV received the support of a fellowship from “La Caixa” Foundation (ID 100010434) with fellowship code LCF/BQ/PI23/11970031 (PI: Ana Escorza). SV also acknowledges funding from national institutions,  participating in the Gaia Multilateral Agreement.
\Ana{DGR acknowledges support from the Spanish Ministry of Science and Innovation (MICINN) with the \emph{Juan de la Cierva} fellowship program under contract JDC2022-049054-I. DGR acknowledges support from the Spanish Ministry of Science and Innovation (MICINN) with the grant No. PID2023-149439NB-C41. DGR acknowledges support from the Spanish Ministry of Science and Innovation (MICINN) with the grant No. PID2023-146453NB-I00 (PLAtoSOnG, PI: Beck).}
\Ana{SS would like to acknowledge the support of Research
Foundation-Flanders (grant number: 1239522N).}
\Ana{LS is FRS-FNRS research director}
MAM acknowledges support from the ``La Caixa'' Foundation (ID 100010434) under the fellowship code LCF/BQ/PI23/11970035.
DAGH acknowledges the support from the State Research Agency (AEI) of the Ministry of Science, Innovation and Universities (MICIU) of the Government of Spain, and the European Regional Development Fund (ERDF), under grant PID2023-147325NB-I00/AEI/10.13039/501100011033. 
Based on observations collected at the European Southern Observatory under ESO programme 111.24UZ (PI: Ana Escorza).
This article is based on observations made in the Observatorios de Canarias del IAC with the HERMES spectrograph, mounted on the Mercator Telescope and operated on the island of La Palma in the Observatorio del Roque de los Muchachos (CAT program 55-Mercator4/24B; PI: Ana Escorza). The HERMES spectrograph is supported by the Fund for Scientific Research of Flanders (FWO), Belgium, the Research Council of KU Leuven, Belgium, the Fonds National de la Recherche Scientifique (F.R.S.-FNRS), Belgium, the Royal Observatory of Belgium, the Observatoire de Genève, Switzerland and the Thüringer Landessternwarte Tautenburg, Germany.
This work used the Fourth Data Release of the GALAH Survey \citep{2025Buder}. The GALAH Survey is based on data acquired through the Australian Astronomical Observatory, under programs: A/2013B/13 (The GALAH pilot survey); A/2014A/25, A/2015A/19, A2017A/18 (The GALAH survey phase 1); A2018A/18 (Open clusters with HERMES); A2019A/1 (Hierarchical star formation in Ori OB1); A2019A/15, A/2020B/23, R/2022B/5, R/2023A/4, R2023B/5 (The GALAH survey phase 2); A/2015B/19, A/2016A/22, A/2016B/10, A/2017B/16, A/2018B/15 (The HERMES-TESS program); A/2015A/3, A/2015B/1, A/2015B/19, A/2016A/22, A/2016B/12, A/2017A/14, A/2020B/14 (The HERMES K2-follow-up program); R/2022B/02 and A/2023A/09 (Combining asteroseismology and spectroscopy in K2); A/2023A/8 (Resolving the chemical fingerprints of Milky Way mergers); and A/2023B/4 (s-process variations in southern globular clusters). We acknowledge the traditional owners of the land on which the AAT stands, the Gamilaraay people, and pay our respects to elders past and present. This paper includes data that has been provided by AAO Data Central (\url{datacentral.org.au}).
This work has used data from the European Space Agency (ESA) mission Gaia (\url{https://www.cosmos.esa.int/gaia}), processed by the Gaia Data Processing and Analysis Consortium (DPAC, \url{https://www.cosmos.esa.int/web/gaia/dpac/consortium}). Funding for the DPAC has been provided by national institutions, in particular the institutions participating in the Gaia Multilateral Agreement.
Given the observational efforts that support this work, the authors want to thank the support astronomers across all the different observatories involved for their work.
\end{acknowledgements}

\bibliographystyle{aa} 
\bibliography{references}

\clearpage
\begin{appendix}
\onecolumn
\section{HR sample: coverage of the GALAH DR4 parameter space}\label{appendix:Fig1DR4}

\begin{figure}[h]
    \centering
    \includegraphics[width=0.81\textwidth]{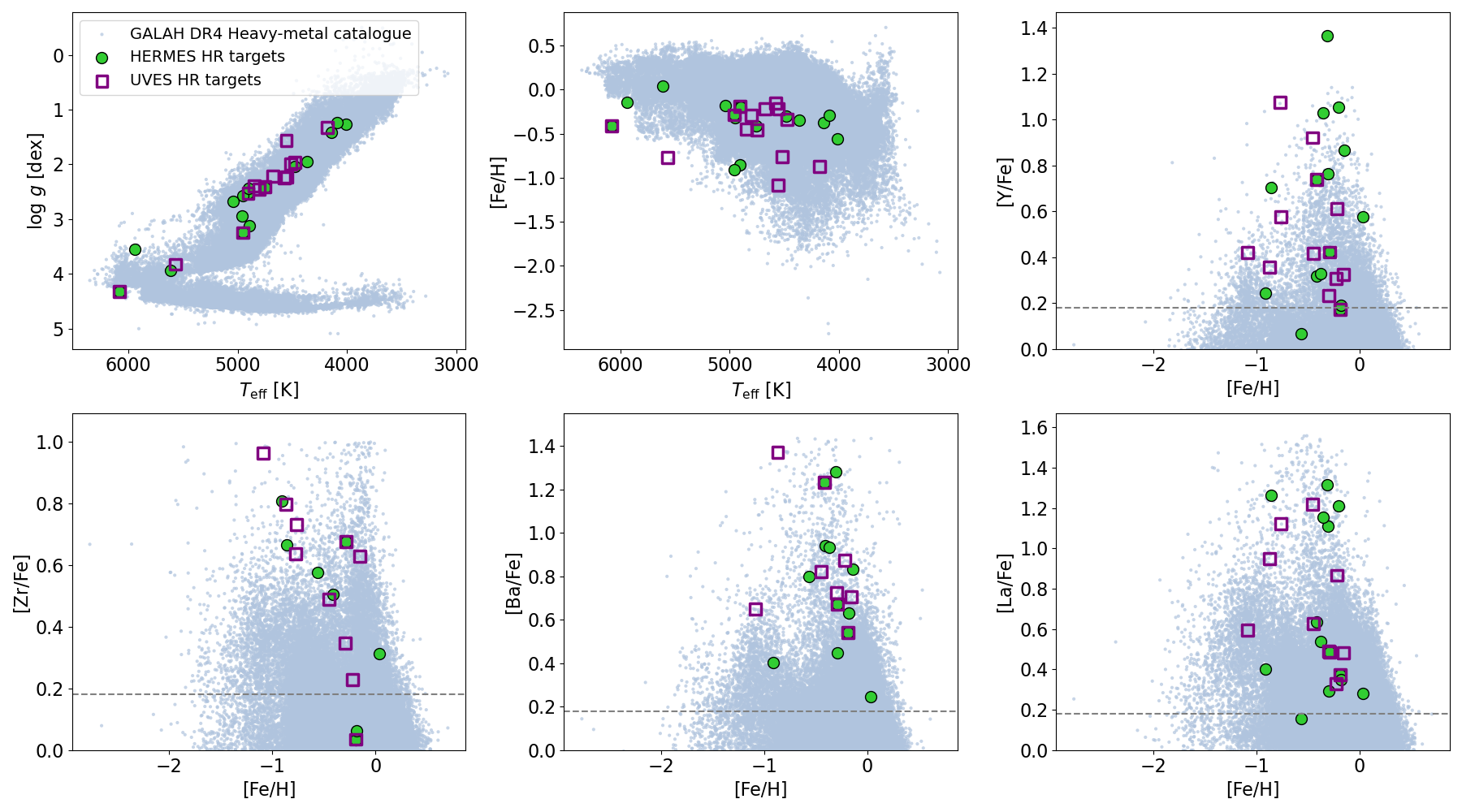}
    \caption{Same as Fig.~\ref{Fig:HRtargets} plotted over the GALAH DR4 “heavy-metal catalogue”}
    \label{Fig:Fig1DR4}
\end{figure}

\section{HR sample: information and observations used in the HR analysis}
\begin{table}[h]
\caption{General information for the targets in the HR sample and details on the observations obtained. }
\label{table:HRtargets}
\centering
\begin{small}
\begin{tabular}{lccccccc}
\toprule
Star ID & RA & Dec & Vmag & Instrument & Exp. t [s] & N & S/N\\
\midrule
BD-00\,1784  &07 40 34.65   &-01 10 38.93   &9.04   &HERMES &1200&1&98@550nm\\
             &              &               &       &UVES   &1500&1&68@390nm/217@564nm\\
BD-01\,2730  &12 50 16.46   &-02 11 58.35   &10.11  &HERMES &3000&1&97@550nm\\
BD-03\,2082  &07 49 46.27   &-03 25 02.49   &9.37   &UVES   &76&1&68@390nm/408@564nm\\
BD-03\,3480  &13 29 21.95   &-04 33 14.14   &10.959 &HERMES &3600&2&98@550nm\\
BD-17\,4712  &17 07 04.43   &-17 50 35.91   &10.29  &UVES   &3000&1&60@390nm/387@564nm\\
CD-25\,3581  &06 45 33.88   &-25 42 31.67   &9.58   &UVES   &3000&1&53@390nm/393@564nm\\
CD-43\,12242 &18 04 27.70   &-43 20 29.96   &9.82   &UVES   &3000&1&95@390nm/465@564nm\\
CD-46\,7169  &11 33 28.14   &-47 21 09.40   &9.97   &UVES   &3000&1&31@390nm/148@564nm\\
HD\,118367   &13 37 26.75   &-46 33 30.44   &8.97   &UVES   &1500&1&117@390nm/538@564nm\\
HD\,170752   &18 33 25.42   &-47 29 43.23   &9.22   &UVES   &1500&1&42@390nm/357@564nm\\
HD\,25584    &04 04 12.98   &+15 01 36.69   &8.03   &HERMES &600&1&127@550nm\\
             &              &               &       &UVES   &500&1&47@390nm/286@564nm\\
HD\,285405   &04 03 48.42   &+15 51 27.20   &11.79  &HERMES &3600&3&108@550nm\\
HD\,285923   &04 31 36.15   &+29 55 37.47   &8.42   &HERMES &3600&2&122@550nm\\
HD\,285933   &04 34 37.38   &+15 02 34.87   &10.01  &HERMES &2700&1&93@550nm\\
HD\,295095   &06 26 03.69   &-04 25 05.68   &9.55   &HERMES &1700&2&80@550nm\\
             &              &               &       &UVES   &1500&1&85@390nm/341@564nm\\
HD\,315809   &17 27 01.41   &-29 13 57.40   &10.32  &UVES   &3000&1&121@390nm/308@564nm\\
HD\,325083   &18 13 14.11   &-38 19 37.67   &9.763  &UVES   &3000&1&78@390nm/497@564nm\\
TYC\,169-2026-1 &07 24 45.85&+03 25 12.38   &10.90  &HERMES &3600&1&62@550nm\\
TYC\,179-2074-1 &07 43 00.29&+00 24 33.04   &10.30  &HERMES &3600&1&46@550nm\\
TYC\,808-1232-1 &08 28 45.50&+13 43 52.48   &9.96   &HERMES &2700&1&38@550nm\\
TYC\,5942-514-1 &06 18 09.75&-18 47 03.90   &10.99  &HERMES &3600&2&61@550nm\\
TYC\,6190-434-1 &15 48 17.85&-17 44 15.71   &10.49  &HERMES &3600&1&63@550nm\\
TYC\,8367-1842-1&18 33 41.38&-48 56 35.69   &9.820  &UVES   &3000&1&65@390nm/335@564nm\\
TYC\,8733-988-1 &17 39 56.63&-55 42 54.27   &10.08  &UVES   &3000&1&54@390nm/395@564nm\\
\bottomrule
\end{tabular}
\tablefoot{
The coordinates come from the \textit{Gaia} EDR3 catalogue \citep{GaiaEDR3summary} and the V magnitudes from the Tycho-2 catalogue \citep{Hog2000}, the APASS Landolt-Sloan BVgri photometric catalogue \citep{Munari2014}, or the Fourth US Naval Observatory CCD Astrograph Catalogue (UCAC4; \citealt{Zacharias2013}). The exposure times, number of spectra (N), and  final S/N values, refer to the spectra used (and coadded when N>1) for the spectroscopic analysis (Sects. \ref{sec:HRsample} and \ref{sec:specanalysis}), not including the binary monitoring programme (Sect. \ref{sec:binarity}).}
\end{small}
\end{table}

\twocolumn

\section{Impact of parameter uncertainties on the derived abundances}\label{appendix:uncert}
\Ana{As seen in Sect.~\ref{sec:comparison-params}, our derived surface gravities are on average $\sim 0.2$ dex smaller than those derived by GALAH DR4 and the most discrepant stars reach $\sim -0.4$ dex. Since most s-process element abundances are derived from ionised lines, which are more sensitive to surface gravity \citep[e.g.][]{2012Bergemann}, these offsets may impact the inferred abundances. To ensure that the abundance thresholds for heavy elements proposed in this work are not primarily driven by differences in surface gravity, we performed a series of tests that we describe here. We repeated our abundance determination three times following the methodology described in Sect.~\ref{sec:abundances}, but fixing the $\log g$ value each time to our derived value plus or minus an offset. We used $-0.2$, $+0.2$, and $+0.4$ dex, the latter chosen to encompass the largest offsets found in comparison with GALAH, while keeping all other atmospheric parameters fixed to those derived from the spectral analysis.}

\Ana{From these tests, we find that variations of $\pm 0.2$ dex in $\log g$ have a negligible impact ($\sim 0.01$–$0.1$ dex) on the derived abundances, lying always within the estimated abundance uncertainties. However, adopting a larger offset of $+0.4$ dex results in a more significant effect, with average abundance differences of $\sim +0.09$ and $+0.12$ dex for \ion{Y}{II}, \ion{Zr}{II}, and \ion{La}{II}, and a smaller variation of $\sim 0.009$ dex for \ion{Ba}{II}. All these variations are, however, smaller than the threshold suggested in Sect.~\ref{sec:bastars}, giving us confidence that even in extreme cases, where the adopted atmospheric parameters differ significantly, we can reliably propose an abundance threshold that truly reflects the s-process composition.}
\end{appendix}

\end{document}